\documentclass[pra,twocolumn,showpacs,preprintnumbers,amsmath,amssymb]{revtex4}

\usepackage{graphicx}
\usepackage{dcolumn}
\usepackage{bm}

\begin{document}

\preprint{APS/123-QED}

\title{Long-Distance Quantum Communication with Neutral Atoms}

\author{Mohsen Razavi}
\email{mora158@mit.edu}
\author{Jeffrey H. Shapiro}
 \affiliation{Research Laboratory of Electronics\\
Massachusetts Institute of Technology\\
Cambridge, Massachusetts 02139 USA}

\date{\today}

\begin{abstract}
The architecture proposed by Duan, Lukin, Cirac, and Zoller (DLCZ) for long-distance quantum communication with atomic ensembles is analyzed.  Its fidelity and throughput in entanglement distribution, entanglement swapping, and quantum teleportation is derived within a framework that accounts for multiple excitations in the ensembles as well as loss and asymmetries in the channel.  The DLCZ performance metrics that are obtained are compared to the corresponding results for the trapped-atom quantum communication architecture that has been proposed by a team from the Massachusetts Institute of Technology and Northwestern University (MIT/NU).  Both systems are found to be capable of high-fidelity entanglement distribution.  However, the DLCZ scheme only provides conditional teleportation and repeater operation, whereas the MIT/NU architecture affords full Bell-state measurements on its trapped atoms.   Moreover, it is shown that achieving unity conditional fidelity in DLCZ teleportation and repeater operation requires ideal photon-number resolving detectors.  The maximum conditional fidelities for DLCZ teleportation and repeater operation that can be realized with non-resolving detectors are 1/2 and 2/3, respectively.
\end{abstract}


\maketitle

\section{INTRODUCTION}
\label{sect:intro}  
Quantum information science is an emerging discipline whose theoretical promise---for revolutionary advances in secure communications, precision measurements, and computational power---has far outstripped its experimental achievements to date.  Networked applications of quantum information processing, see e.g. \cite{distqp}, may provide an excellent route for the initial deployment and attendant continuing development of this new technology.  For these applications, few-qubit processors of relatively modest fidelity that are connected by similarly capable teleportation links \cite{Bennett} will suffice. This kind of quantum information processing  must be built on reliable means for transforming flying qubits into standing qubits, so that entanglement can be established and maintained between systems that are separated by long distances.   Photons are the only feasible flying qubits for long-distance transmission, and the hyperfine levels of neutral atoms provide attractive venues for standing qubits.  To date a variety of schemes have been suggested for photon-mediated, neutral-atom quantum communication \cite{qcomm1}--\cite{qcomm4}---using either trapped atoms in cavity quantum electrodynamics (cavity-QED) setups or atomic ensembles---and experimental progress has been made toward realizing elements of these architectures \cite{exp1}--\cite{exp10}. Nevertheless, much experimental work needs to be done before any of these systems could demonstrate the long-distance qubit teleportation and few-qubit processing that will be needed for networked applications.  Moreover, in advance of any such experimental progress it will be valuable to understand the quantum communication performance---throughput in entanglement distribution and fidelity of qubit teleportation---that can be achieved in these architectures.  This paper will address such performance questions for the atomic ensemble scheme of Duan, Lukin, Cirac, and Zoller (termed DLCZ hereafter)
  \cite{qcomm4,3D}, and compare the results thus obtained with those previously derived \cite{qcomm2,Brent} for the trapped-atom architecture suggested by a team from the Massachusetts Institute of Technology and Northwestern University (termed MIT/NU hereafter) \cite{trap}.  

Entanglement is the fundamental resource for quantum communication, hence entanglement distribution is the initial task to be completed by a quantum communication system.  Putting aside the fact that the MIT/NU architecture employs single trapped atoms for its quantum memories (QMs), whereas the DLCZ architecture uses atomic ensembles for its QMs, there is a more abstract way to distinguish between their respective approaches to entanglement distribution.  MIT/NU entanglement distribution can be termed a to-the-memory architecture.  As shown in Fig.~1(a), to-the-memory entanglement-distribution first produces a pair of entangled photons from an optical source, then lets them propagate to remote locations for capture and storage in a pair of quantum memories.  DLCZ entanglement is a from-the-memory approach, see Fig.~1(b), which relies on entanglement swapping \cite{swap}.  Here, entanglement is established between a memory qubit and a photon at location $A$ and similarly for another memory-photon qubit pair at location $B$.  The photons then propagate to the midpoint between $A$ and $B$ where a Bell-state measurement (BSM) annihilates them, leaving the memory qubits at $A$ and $B$ in an entangled state.  
\begin{figure}
\centering
\includegraphics [width=3.25in] {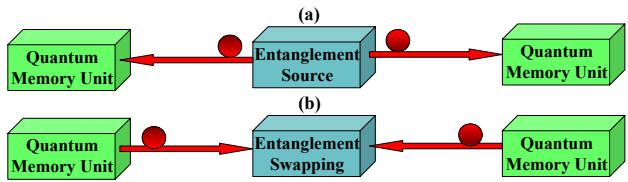}
\caption{\label{telepapp}
(Color online) Two architectures for entanglement distribution: (a) to-the-memory distribution, in which two entangled photons propagate to and are loaded into quantum memories; (b) from-the-memory distribution, in which entangled memory-photon qubit pairs are created, the photons propagate to a common location where a Bell-state measurement annihilates them, leaving the memories in an entangled state.}
\end{figure}

From-the-memory entanglement distribution is accomplished in the DLCZ architecture by weak coherent pumping of a Raman transition in each ensemble followed by path-erasing photodetection.  In particular, collective excitation of an ensemble will radiate a single photon in a well-defined spatial mode.  The output modes from the two ensembles are then routed to a common location, e.g. via optical fibers, combined on a 50/50 beam splitter, and detected by a pair of single-photon counters.  Because the ensembles are coherently pumped, because the probability that \em both\/\rm\ will emit Raman photons will be very low compared to the single-ensemble emission probability, and because the beam-splitter combining erases any which-way information, observation of a photocount on one---and only one---detector heralds the entanglement of the two ensembles.   In contrast, the MIT/NU architecture uses a to-the-memory configuration.  Cavity-enhanced spontaneous parametric downconversion (a dual optical parametric amplifier) is used to generate an ultrabright, narrowband stream of polarization-entangled photon pairs.  One photon from each pair is sent down optical fiber to its own trapped-atom quantum memory.  A non-destructive, cycling-transition procedure is then used to deduce that the two atoms have been loaded, i.e., the memories have absorbed the entangled photon pair.

A fundamental limitation on the entanglement-distribution performance---throughput and fidelity---for both the DLCZ and MIT/NU architectures arises from a common cause: both rely on entangled-Gaussian input states.  For the DLCZ architecture, the input state in question is the joint state of each atomic ensemble and its Stokes-light output.  For the MIT/NU architecture, the entangled-Gaussian input state is that of the signal and idler beams from its dual optical parametric amplifier source. A comprehensive Gaussian-state analysis of the MIT/NU construct has been given in Refs. \cite{qcomm2,Brent} considering various sources of failure in the system. In Sec.~\ref{DLCZEntg} of this paper we develop a similar Gaussian-state theory for DLCZ entanglement distribution, which we compare, in Sec.~\ref{Seccomp}, with the corresponding analysis of the MIT/NU architecture.  In our treatment, we study the effects of pump phase/amplitude mismatch as well as possible asymmetries in the channel/detectors. Then, in Sec.~\ref{Sectel}, we examine the fidelity achieved by the DLCZ repeater and teleportation protocols, under the assumption that successful entanglement distribution has occurred.  

\section{DLCZ entanglement distribution} \label{DLCZEntg}

\begin{figure}
\begin{center}
\includegraphics[width=3in]{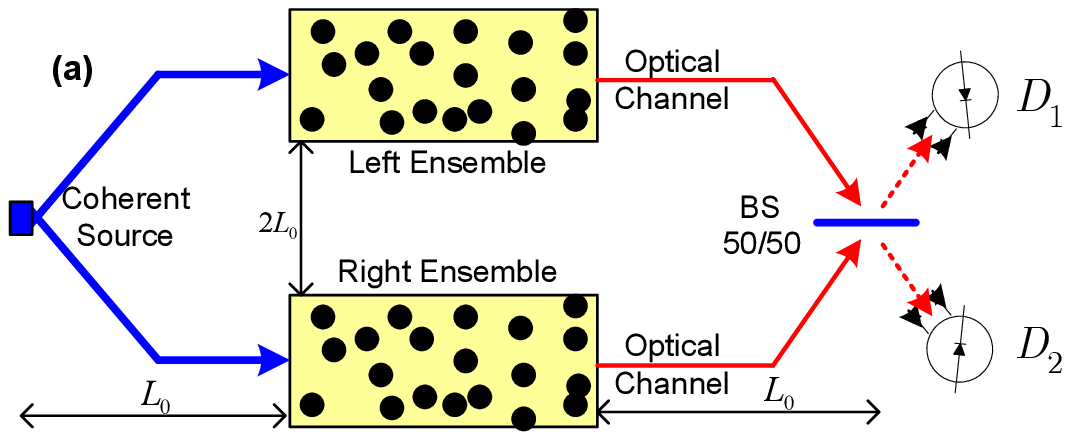}
\end{center}
\begin{center}
\includegraphics[width=1.5in]{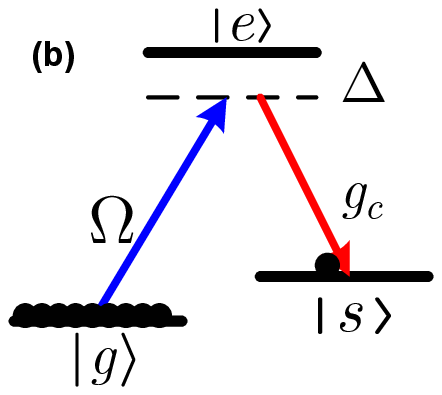} 
\end{center}
\vspace*{-.3in}
 \caption{ \label{Fig1} 
(Color online) (a) DLCZ architecture for entanglement distribution. A coherent laser source, located at the midpoint between two atomic ensembles, induces Raman transitions in these ensembles.   Occurrence of a single click on one---and only one---detector heralds the protocol's success, i.e., the atomic ensembles are then expected to be entangled. (b) $\Lambda$-level structure for the atoms in the ensembles:  $\Omega$ is the Rabi frequency associated with the off-resonant (detuning $\Delta$) pumping of the $|g\rangle\rightarrow |e\rangle$ transition; and $g_c$ is the coupling coefficient for the $|e\rangle\rightarrow |s\rangle$ transition. }
\end{figure}

The DLCZ protocol for entangling two nonlocal atomic ensembles is shown schematically in Fig.~\ref{Fig1}(a).  The two ensembles---each consisting of $N_a $ identical atoms with $\Lambda$-level configurations, as shown in Fig.~\ref{Fig1}(b)---are coherently pumped using a weak, off-resonant laser such that the probability of occurrence of a Raman transition from the ground level $\left| g \right\rangle $ to the metastable level $\left| s \right\rangle $, is very low. Because each atom in the left ($L$) or right ($R$) ensemble is equally likely to undergo a Raman transition, the resulting Raman photon is matched to the symmetric collective atomic mode represented by the operator 
\begin{equation}
S_A = \frac{1}{\sqrt {N_a }} \sum_{n=1}^{N_a} {\left| g \right\rangle _{A_{n\,} A_n}\!\!{\left\langle s \right|}},
\end{equation}
where the sum is over the atoms in ensemble $A$, for $A \in \{L,R\}$. The forward-scattered Stokes light from such a Raman transition in each ensemble is routed over an $L_0$-km-long path to the midpoint between the locations of the two ensembles.  There, the outputs from these optical channels are combined on a 50/50 beam splitter (BS) prior to measurement by a pair of single-photon detectors, $D_1 $ and $D_2 $, whose dark-count rates will be assumed to be negligible.  Assume that the setups for the generation, distribution, and detection of Raman photons are completely symmetric.  Furthermore, suppose that only one ensemble undergoes a Raman transition, and this transition is detected by detector $D_j$ registering a count.   Then, because the pumping is coherent and the beam splitter erases which-path information the two ensembles will be left in the entangled state
\begin{equation}
|\psi_j\rangle \equiv (|0\rangle_L|1\rangle_R  + (-1)^j|1\rangle_L|0\rangle_R)/\sqrt{2},\;
\mbox{ for $j = 1,2,$}
\label{entangle}
\end{equation}
where 
\begin{equation}
|0\rangle_A \equiv \bigotimes_{n=1}^{N_a}|g\rangle_{A_n}\;\mbox{and }\;
|1\rangle_A\equiv S^\dagger_A|0\rangle_A\; \mbox{ for $A = L,R,$}
\end{equation}
are the atomic ground state and symmetric collective excited state of ensemble $A$.  

There are a variety of ways by which DLCZ entanglement distribution can err. First, there is the possibility that more than one Raman transition has occurred, e.g. two transitions in one ensemble or one transition in each.  A single detector click might still occur in this case.  For example, all but one of these multiple Raman photons might be lost en route to the detection setup, or all but one might fail to be detected because of sub-unity detector quantum efficiency.  Alternatively, if the single photon detectors in Fig.~\ref{Fig1}(a) are Geiger-mode avalanche photodiodes (APDs)---which are incapable of distinguishing multiple-photon pulses from single-photon pulses---then the clicking of one and only one of the two detectors would not guarantee that only one Raman photon had been seen.  In all of these circumstances the DLCZ protocol would announce that the ensembles were now entangled, according to Eq.~(\ref{entangle}), when in fact the joint state of these two ensembles would not be given by this expression.  Hence any reliance on Eq.~(\ref{entangle}), say for the performance of qubit teleportation, would be unwarranted.  

Another reason that the DLCZ ensembles might not be left in one of the maximally entangled states from Eq.~(\ref{entangle}) is due to asymmetries in the system. For example, an imbalance between the total loss seen by each Raman photon and/or different pump power at each ensemble will strengthen the $|1 \rangle _ L |0 \rangle _ R$ component of Eq.~(\ref{entangle}) relative to its $|0 \rangle _ L |1 \rangle _ R$ component or \em vice versa\/\rm.  Phase mismatch, arising from different pump phases and/or unequal accumulated phases en route to the detectors, can also severely degrade the fidelity of entanglement distribution. We will assume that the latter source of phase mismatch has been compensated by means of additional phase shifters. To compensate for the former, however, requires achieving perfect phase stability between the laser pump beams that are applied to a pair of widely-separated atomic ensembles. We assume a random pump-phase offset to account for possible errors in this process.  

Other error mechanisms for DLCZ entanglement distribution include detector dark counts, which can masquerade as Raman photon detections, and the spatial-mode mismatch, which arises in a 3D treatment of the atomic ensembles \cite{3D}.  The dark-count rates of silicon Geiger-mode APDs are sufficiently low, at wavelengths of interest for several atomic species, that we shall neglect dark counts in our analysis.  Moreover, we neglect the subtleties that arise in a 3D treatment of the problem by assuming a pencil-shaped ensemble with almost unity Fresnel number (ensemble cross-sectional area divided by the product of its length and the pump-laser's wavelength)  \cite{pencil}. We also neglect the effects of spontaneous emission, whose significance is reduced by the off-resonant pumping and the signal-to-noise ratio enhancement afforded by the collective atomic  behavior \cite{qcomm4}.  Finally, we assume that the coherence time of the ensembles is long enough to allow for a few runs of each protocol in a long-distance scenario \cite{cohtime, Kuz2}.

In what follows, we will derive the performance of DLCZ entanglement distribution when it is limited by the possibility of multiple Raman-transition events.  We start from the Gaussian entangled-state characterization of the atomic ensembles and their associated Stokes light, allowing for pump phase/amplitude errors.   The Stokes light is then propagated through to the detection system, considering propagation losses as well as sub-unity quantum efficiencies. The resulting transformed Gaussian state is then used to evaluate the fidelity and throughput of the DLCZ protocol when we employ either non-resolving photon detectors (NRPDs), i.e., detectors that are incapable of distinguishing single-photon from multiple-photon events, or photon-number resolving detectors (PNRDs) that can draw such distinctions. 

\subsection{Atomic-Photonic Initial Joint State} 

Neglecting spontaneous emission, the joint state of a $\Lambda$-level atomic ensemble---held within a ring cavity of decay rate $\kappa$ and pumped for $t_\Delta$\,sec at Rabi frequency $\Omega$ and detuning $\Delta$---and its associated Stokes light is the entangled (two-mode squeezed) state \cite{qcomm4}:
\begin{equation} \label{(12)}
	|\psi\rangle  = \frac{1}{\cosh r} \sum_{n=0}^{N_a} \frac{(S_a^\dag  a_p^\dag \, e^{i \theta} \tanh r)^n\left| {0_a } \right\rangle \left| {0_p } \right\rangle}{n!}. 
\end{equation}
In Eq.~(\ref{(12)}), $S_a$ and $a_p $ are the annihilation operators for the symmetric collective atomic mode and the effective mode for the Stokes light, respectively, $\theta$ is the pump-phase offset,
and
\begin{equation}
\cosh r = \exp (2N_a \left| {\Omega g_c } \right|^2 t_\Delta  /\Delta ^2 \kappa ),
\end{equation}
specifies the squeeze parameter, $r$, for this state.  Our calculations below will rely on an equivalent specification for this joint state, i.e., its antinormally-ordered characteristic function \cite{Gnoise}:
\begin{eqnarray}
\lefteqn{\chi _A^{\nu \mu } (\zeta _a ,\zeta _p ) \equiv  \left\langle {D_A (S_a,\zeta _a )D_A (a_p ,\zeta _p )} \right\rangle } \nonumber \\  
 &=&  \exp \left[ { - \left| \mu  \right|^2 (\left| {\zeta _a } \right|^2  + \left| {\zeta _p } \right|^2 ) - 2{\mathop{\rm Re}\nolimits} (\mu \nu \zeta _a^ *  \zeta _p^ *  )} \right], \label{(E27)}
\end{eqnarray}
where $\nu  =  - \sinh r \exp(i \theta)$, $\mu  = \cosh r$, and $D_A (a,\zeta ) \equiv e^{ - \zeta ^ *  a} e^{\zeta a^\dag  } $ is the antinormally-ordered displacement operator.  Because $\chi_A^{\nu \mu}$ is a Gaussian form, we say that $|\psi\rangle$ is a Gaussian state.  

Using Eq.~(\ref{(E27)}), we have that the joint state, $\rho_{\rm in }$, of the two atomic ensembles and their Stokes light at the optical channel inputs in Fig.~\ref{Fig1}(a) has the following antinormally-ordered characteristic function:
\begin{equation} \label{(13)}
\chi _A^{\rho _{\rm in} } (\zeta _a^L ,\zeta _a^R ,\zeta _p^L ,\zeta _p^R ) = \chi _A^{\nu _L \mu_L } (\zeta _a^L ,\zeta _p^L )\chi _A^{\nu _R \mu_R } (\zeta _a^R ,\zeta _p^R ),
\end{equation}
where $\nu _A / \mu_A   = \sqrt {p_{c_A}} \exp (i\theta _A ) $ and $A \in \{L,R\} $.  Here, $\theta _L $ and $\theta _R $ model the pump-phase offsets for the left and right ensembles, respectively.  Because of the short-duration Raman pumping employed in the DLCZ protocol, making these time-independent phase shifts into random variables---as we will do later---can account for imperfect coherence in the pumping of the two atomic ensembles. From Eq.~(\ref{(12)}), the probability of exciting a single Raman transition in ensemble $A$ is $p_{c_A}(1-p_{c_A})$, which becomes $p_{c_A} \ll 1$ under weak pumping conditions. 

\subsection{Optical Channel Output} 

\begin{figure}
   \centering
      \includegraphics[width=3.25in]{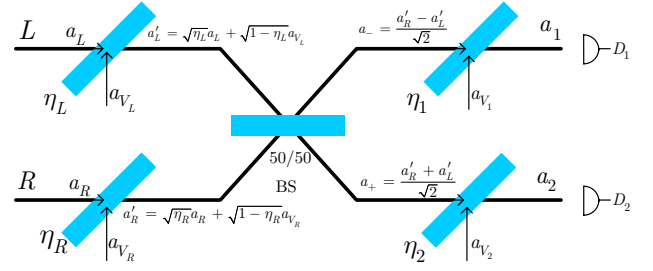}
   \caption
   { \label{FA1} 
(Color online) Notional model for the optical channels shown in Fig.~\ref{Fig1}(a).   Fictitious beam splitters are used to account for the loss of Raman photons and the quantum noise introduced by propagation from the atomic ensembles to the 50/50 beam splitter in Fig.~\ref{Fig1}(a), and by the sub-unity quantum efficiencies of the detectors appearing in that figure.  The detectors in Fig.~\ref{FA1} have unity quantum efficiencies.}
   \end{figure}

Figure~\ref{FA1} depicts our model for the optical channels shown in Fig.~\ref{Fig1}(a).  Here, propagation losses between the atomic ensembles and the 50/50 coupling beam splitter from Fig.~\ref{Fig1}(a) are represented by fictitious beam splitters whose free input ports inject vacuum-state quantum noise.  Additional fictitious beam splitters are placed after the 50/50 coupling beam splitter---again with vacuum-state quantum noise injected through their free input ports---to account for the sub-unity quantum efficiencies of the detectors shown in Fig.~\ref{Fig1}(a).  Thus, detectors $D_1$ and $D_2$ in Fig.~\ref{FA1} are taken to have unity quantum efficiencies.  The transmissivity, vacuum field, and output field associated with each beam splitter have been shown in the figure. It can be seen that the optical channel consists of linear optical elements for which we can write input-output operator relations.  Doing that, we then have that the annihilation operators for the fields reaching the Fig.~\ref{FA1} detectors are \cite{Teich}
\begin{subequations}
\label{(B1)}
\begin{eqnarray}
\lefteqn{a_1 =   \sqrt {\frac{\displaystyle \eta}{\displaystyle 2}} (\sqrt {\eta _R } a_R  + \sqrt {1 - \eta _R } a_{V_R } )} \nonumber\\
&\,& -\sqrt {\frac{\displaystyle \eta}{\displaystyle 2}} (\sqrt {\eta _L } a_L  + \sqrt {1 - \eta _L } a_{V_L } )
+ \sqrt {1 - \eta _1 } a_{V_1 } \\
\lefteqn{a_2  =  \sqrt {\frac{\displaystyle \eta}{\displaystyle 2}}  (\sqrt {\eta _R } a_R  + \sqrt {1 - \eta _R } a_{V_R } ) }\nonumber \\ 
 &\,&+\sqrt {\frac{\displaystyle \eta}{\displaystyle 2}}  (\sqrt {\eta _L } a_L  + \sqrt {1 - \eta _L } a_{V_L } )
 + \sqrt {1 - \eta _2 } a_{V_2 },
\end{eqnarray}
\end{subequations}
where $a_{V_{L,R}} $ and $a_{V_{1,2}} $ are in their vacuum states, and $a_L$ and $a_R$ are, respectively, the associated field operators for the Raman photons originating from the left and right ensembles. These linear transformations preserve the Gaussian-state nature of their inputs.   In particular, using Eq.~(\ref{(B1)}), we have that the joint state, $\rho_{\rm out}$, of the two atomic ensembles and their Stokes light arriving at the Fig.~\ref{FA1} detectors has an antinormally-ordered characteristic function given by
\begin{eqnarray}
\lefteqn{\chi _A^{\rho _{\rm out} } (\zeta _a^L ,\zeta _a^R ,\zeta _{p1} ,\zeta _{p2} )  }
\nonumber\\ 
&\equiv&  \left\langle {D_A (S_L ,\zeta _a^L )D_A (S_R ,\zeta _a^R )D_A (a_1 ,\zeta _{p1} )D_A (a_2 ,\zeta _{p2} )} \right\rangle \nonumber \\
& = & \chi _A^{\rho _{\rm in} } (\zeta _a^L ,\zeta _a^R ,\sqrt {\eta _L  } \zeta _p^ -  ,\sqrt {\eta _R  } \zeta _p^ +  )   \nonumber \\
&\,& \times  \exp \left[  - (1 - \eta _1 ) \left| {\zeta _{p1}  } \right|^2 - (1 - \eta _2 ) \left| {\zeta _{p2}  } \right|^2   \right] \nonumber \\
&\,& \times  \exp \left[  - (1 - \eta _L ) \left| {\zeta _p^-  } \right|^2 - (1 - \eta _R ) \left| {\zeta _p^+  } \right|^2   \right]
 , \label{(B4)}
\end{eqnarray}
where
\begin{equation}
\label{zetapm}
\zeta _p^ \pm   = \sqrt {\frac{\displaystyle \eta}{\displaystyle 2}} \zeta _{p2}  \pm \sqrt {\frac{\displaystyle \eta}{\displaystyle 2}}\zeta _{p1} \, .  
\end{equation}
Then, by employing Eqs.~(\ref{(E27)}) and (\ref{(13)}) in Eq.~(\ref{(B4)}), we get
\begin{eqnarray}
\lefteqn{\chi _A^{\rho _{\rm out} } (\zeta _a^L ,\zeta _a^R ,\zeta _{p1} ,\zeta _{p2} ) = 
\exp \left[ 
- \frac{\alpha_L}{2} \left| {\zeta _a^L } \right|^2\right. }
\nonumber \\   && 
- \frac{\beta_L }{2} \left| {\zeta _p^- } \right|^2
 -  \gamma_L {\mathop{\rm Re}\nolimits} \{ e^{i\theta _L } {\zeta _a^L} ^ *  {\zeta _p^ -}  ^ *  \}  
  \nonumber \\
 &&  - \delta {\mathop{\rm Re}\nolimits} \{ {\zeta _p^+}   {\zeta _p^ -}  ^ *  \}
 -  \gamma_R {\mathop{\rm Re}\nolimits} \{ e^{i\theta _R } {\zeta _a^R} ^ *  {\zeta _p^ +}  ^ *  \}
- \frac{\alpha_R}{2} \left| {\zeta _a^R } \right|^2 \nonumber \\
&& \left. - \frac{\beta_R }{2} \left| {\zeta _p^+ } \right|^2  \right] ,
\label{(C1)}
\end{eqnarray}
where
\begin{subequations}
\begin{eqnarray}
& \alpha_A  =  2 |\mu_A|^2 = 2 / (1 - p_{c_A}) \, , & \\
& \beta_A = \eta_A p_{c_A} \alpha_A  +  (\eta_1 + \eta_2)/(\eta_1  \eta_2) \, , & \\
& \gamma_A  =  \sqrt{\eta_A p_{c_A}} \alpha_A \, , & \\
& \delta  =  (\eta_1 - \eta_2)/(\eta_1  \eta_2) \, , &
\end{eqnarray}
\end{subequations}
for $A = L,R$. Therefore, we can think of the complex vector $[\zeta _a^L ,\zeta _p^ - , \zeta _p^ + , \zeta _a^R  ]$ as a zero-mean Gaussian random vector whose covariance matrix can be determined from Eq.~(\ref{(C1)}), see appendix for details. In our subsequent analysis we will use this fact to evaluate probabilities of interest via Gaussian moment relations. 

The output density operator can be written in terms of its respective antinormally-ordered characteristic function via the following operator-valued inverse Fourier transform relation:
\begin{eqnarray}
\label{(B5)}
\rho _{\rm out} & = &\int\!{\frac{{ {\rm d}^2 \zeta _a^L }}{\pi }\int\!{\frac{{ {\rm d}^2 \zeta _a^R }}{\pi } D_N (S_L ,\zeta _a^L )D_N (S_R ,\zeta _a^R )}} \nonumber \\
& {\rm \!} & \times
\int\! \frac{{ {\rm d}^2 \zeta _{p1} }}{\pi } \int\! {\frac{{{\rm d}^2 \zeta _{p2} }}{\pi }\,\chi _A^{\rho _{\rm out} } (\zeta _a^L ,\zeta _a^R ,\zeta _{p1} ,\zeta _{p2} )}
\nonumber \\ 
&{\rm \!}& \times D_N (a_1 ,\zeta _{p1} )D_N (a_2 ,\zeta _{p2} )  \, , 
\end{eqnarray}
where $D_N (a,\zeta ) \equiv e^{ - \zeta a^\dag  } e^{\zeta ^ *  a} $ is the normally-ordered displacement operator, and $\int\!{{\rm d}^2 \zeta} \equiv \int{\int\!{{\rm d} \zeta_r{\rm d} \zeta_i}}$, where $\zeta_r$ and $\zeta_i$ are, respectively, the real and imaginary parts of $\zeta$. We use this convention throughout the paper.

\subsection{Measurement Modules}

The occurrence of a detection click on one, and only one, of the photodetectors $D_1 $ and $D_2 $ is used to herald entanglement distribution in the DLCZ protocol.  We shall consider both non-resolving single-photon detectors (NRPDs), which are incapable of distinguishing multiple-photon pulses from single-photon pulses, as well as photon-number resolving detectors (PNRDs), which are capable of making such distinctions.  The latter, which were not considered in the original DLCZ protocol, allow suppression of  error events that were undetectable with NRPDs, i.e., the PNRD version of the entanglement-distribution protocol heralds entanglement distribution when exactly one photon is detected by the $\{D_1,D_2\}$ pair.   

Let $M_1 $ and $M_2$ be measurement projectors on the joint state space of the $a_1$ and $a_2$ modes that represent DLCZ heralding events in which detections occur on $D_1$ and $D_2$, respectively.  For example, $M_1$, in the NRPD case, implies the detection of a single click (one or more photons) on detector $D_1$ and none on detector $D_2$; in the PNRD case this operator implies the detection of exactly one photon on $D_1$ and none on $D_2$.   From these descriptions we get the following explicit forms for $M_1$ and $M_2$:
\begin{eqnarray}
M_1 &=& \left\{\begin{array}{ll}
\left| 1 \right\rangle _{1\,1} \!\left\langle 1 \right| \otimes \left| 0 \right\rangle _{2\,2}\!\left\langle 0 \right|,   &
\mbox{PNRD}\\[.1in]
(I_1  - \left| 0 \right\rangle _{1\,1}\!\left\langle 0 \right|) \otimes \left| 0 \right\rangle _{2\,2}\!\left\langle 0 \right| , 
& \mbox{NRPD,}\end{array}\right. 
\label{M1}\\[.1in]
M_2 &=& \left\{\begin{array}{ll}
\left| 0 \right\rangle _{1\,1}\!\left\langle 0 \right| \otimes \left| 1 \right\rangle _{2\,2}\!\left\langle 1 \right| ,
& \mbox{PNRD}\\[.1in]
\left| 0 \right\rangle _{1\,1}\!\left\langle 0 \right| \otimes (I_2  - \left| 0 \right\rangle _{2\,2}\!\left\langle 0 \right|),
& \mbox{NRPD,}\end{array}\right.
\label{M2}
\end{eqnarray}
where $I_1 $ and $I_2 $ denote the identity operators for the $a_1$ and $a_2$ modes, respectively.

Suppose that the DLCZ protocol (with either NRPDs or PNRDs) has heralded entanglement distribution, based on observing a click from $D_j$ and no click from $D_i$, where $i, j = 1,2$ and $i\neq j$.   The post-measurement joint density operator for the two atomic ensembles, $\rho _{{\rm pm}_j}$, can be found by projecting with $M_j$, tracing out the photonic variables, and renormalizing, viz.
\begin{equation}
\label{(B6)}
 \rho _{{\rm pm}_j}  = \frac{{{\rm tr}_{1,2} (\rho _{\rm out} M_j  )}}{{P_j }},
\end{equation}
where
\begin{equation}
\label{(B10)}
P_j  =   {\rm tr}( {\rho _{\rm out} M_j })
\end{equation}
is the probability that the conditioning event $M_j$ has occurred. The total probability that the DLCZ protocol heralds an entanglement distribution is then $P_{\rm herald}  = P_1  + P_2 $. Note that $P_{\rm herald}$ is \em not\/\rm\ the probability that the atomic ensembles have been placed into the entangled state $|\psi_j\rangle$ if $M_j$ has occurred. The success probability, $P_{\rm success} $, for creating this entanglement is 
\begin{equation}
P_{\rm success} = P_1\langle \psi_1|\rho_{{\rm pm}_1}|\psi_1\rangle 
+ P_2\langle \psi_2|\rho_{{\rm pm}_2}|\psi_2\rangle,
\end{equation}
i.e., the heralding probabilities, $P_j$,  must be multiplied by their associated fidelities,
$F_j \equiv \langle\psi_j|\rho_{{\rm pm}_j}|\psi_j\rangle$, for successful entanglement distribution.  These fidelities will be less than unity, because of higher-order (multiple-photon) components in the input state $\rho_{\rm in}$.    

In the remainder of this section, we shall find the post-measurement states, $\{\rho_{{\rm pm}_j}\}$, the heralding probabilities, $\{P_j\}$, and the fidelities of entanglement, $\{F_j\}$, for DLCZ entanglement distribution.  Both PNRD and NRPD systems will be considered.   

\subsubsection{Photon-Number Resolving Detectors}
It can be easily verified that for any single-mode annihilation operator $a$ and complex variable $\zeta $, we have
\begin{equation}
\label{(E13)}
\langle 0 |D_N (a,\zeta )| 0 \rangle  = 1 \mbox{ and }   \langle 1 |D_N (a,\zeta )| 1 \rangle  = 1 - \left| \zeta  \right|^2  .
\end{equation}
Using these results, together with Eqs.~(\ref{(B5)}) and (\ref{(B6)}) plus the PNRD cases from Eqs.~(\ref{M1}) and (\ref{M2}), we get
\begin{eqnarray}
\label{(B8)}
\lefteqn{\hspace*{-.1in}\rho _{{\rm pm}_j}  =  \frac{1}{{P_j }}\int \!{\frac{{{\rm d}^2 \zeta _a^L }}{\pi }\int \!{\frac{{{\rm d}^2 \zeta _a^R }}{\pi }} D_N (S_L ,\zeta _a^L )D_N (S_R ,\zeta _a^R ) }\nonumber }\\
&\,&\times \int \!\frac{{{\rm d}^2 \zeta _{p1} }}{\pi }\int \!{\frac{{{\rm d}^2 \zeta _{p2} }}{\pi } } \,\chi _A^{\rho _{\rm out} } (\zeta _a^L ,\zeta _a^R ,\zeta _{p1} ,\zeta _{p2}) 
\nonumber \\ 
&\,& \times \left( {1 - \left| {\zeta _{pj} } \right|^2 } \right),
\end{eqnarray}
whence, by means of Eq.~(\ref{(B10)}) and the identity ${\rm tr}(D_N (a,\zeta )) = \pi \delta (\zeta )$, 
\begin{equation}
\label{(B9)}
P_j  = \int\! {\frac{{{\rm d}^2 \zeta _{p1} }}{\pi }\int\! {\frac{{{\rm d}^2 \zeta _{p2} }}{\pi }\,\chi _A^{\rho _{\rm out} } (0,0,\zeta _{p1} ,\zeta _{p2} )\left( {1 - \left| {\zeta _{pj} } \right|^2 } \right)} }.
\end{equation}
The above integral can be evaluated from moments that are directly identifiable from the Gaussian characteristic function in Eq.~(\ref{(C1)}), and we obtain (see appendix for details)
\begin{eqnarray}
\label{(15)}
P_j &=& \frac {4}{\eta_1 \eta_2(\beta_L \beta_R - \delta ^ 2)}\nonumber \\
&\,& \times \left( 1- \frac {\beta_L + \beta_R - 2 (-1)^j \delta}{\eta_j(\beta_L \beta_R - \delta ^ 2)} \right) ,  \mbox{for $j =1 ,2$.}
\end{eqnarray}
 
In the special case of a symmetric setup, in which $\eta_L = \eta_R$, $\eta_1 = \eta_2$, $\theta_L = \theta_R$, and $p_{c_L} = p_{c_R} \equiv p_c$, the preceding expression reduces to 
\begin{equation}
\label{(18)}
P_j  = \frac{{(1 - p_c )^2 \eta _s p_c }}{{(\eta _s p_c  + 1 - p_c )^3 }},\quad\mbox{for $j =1 ,2$,}
\end{equation}
where $\eta_s = \eta_L \eta_1$ is the system efficiency.  In this case $P_1 = P_2$ holds, owing to the symmetry of the optical channels and the measurement modules.  More generally, $\eta_1 = \eta_2$ implies $P_1 = P_2$, because this condition suffices to make $D_1$ and $D_2$ photon detections equally likely.  

\subsubsection{Non-Resolving Photon Detectors}
Similar to the PNRD case, we start from 
\begin{equation}
\label{(E16)}
{\rm tr}\!\left[ D_N (a,\zeta )\left( I - \left| 0 \right\rangle \left\langle 0 \right| \right) \right] = \pi \delta (\zeta ) - 1
\end{equation}
along with Eqs.~(\ref{(E13)}), (\ref{(B5)}), (\ref{(B6)}) plus the NRPD cases from Eqs.~(\ref{M1}) and (\ref{M2}), and obtain
\begin{eqnarray}
\label{(B14)}
\rho _{{\rm pm}_j} & = & \frac{1}{{P_j }}\int\!{\frac{{{\rm d}^2 \zeta _a^L }}{\pi }\int\!{\frac{{{\rm d}^2 \zeta _a^R }}{\pi }} \,D_N (S_L ,\zeta _a^L )D_N (S_R ,\zeta _a^R )} \nonumber \\
& \! & \, \, \times \int\!\frac{{{\rm d}^2 \zeta _{p1} }}{\pi }\int\! {\frac{{{\rm d}^2 \zeta _{p2} }}{\pi } }\, \chi _A^{\rho _{\rm out} } (\zeta _a^L ,\zeta _a^R ,\zeta _{p1} ,\zeta _{p2} )
\nonumber \\ 
& \! & \, \, \times \left( {\pi \delta (\zeta _{pj} ) - 1} \right) ,
\end{eqnarray}
where
\begin{eqnarray}
\label{(19)}
P_j  &=&  \int\!{\frac{{{\rm d}^2 \zeta _{p1} }}{\pi }\int\! {\frac{{{\rm d}^2 \zeta _{p2} }}{\pi }\,\chi _A^{\rho _{\rm out} } (0,0,\zeta _{p1} ,\zeta _{p2} )\left( {\pi \delta (\zeta _{pj} ) - 1} \right)} } \nonumber \\
&=& \frac {4} {\eta_{i} (\beta_L + \beta_R - 2 (-1)^j \delta) } \nonumber \\ 
&\,& - \frac {4} {\eta_1 \eta_2 ( \beta_L \beta_R - \delta^2) } ,\; \mbox{ for $i,j =1 ,2$, $i\neq j$}
\end{eqnarray}
For the symmetric setup, the above probability simplifies to
\begin{equation} 
\label{(21)}
P_j =  \frac{{(1 - p_c )\eta _s p_c }}{{(\eta _s p_c  + 1 - p_c )^2 }},\quad\mbox{for $j =1 ,2$}.
\end{equation}
As was the case for PNRDs, $\eta_1 = \eta_2$ is again enough to ensure that $P_1 = P_2$.  Comparison of Eqs.~(\ref{(18)}) and (\ref{(21)}) reveals that $P_j$ for the  NRPD case is higher than $P_j$ for the PNRD case.  This is to be expected, because the heralding events included in the latter probability are a proper subset of those included  in the former.  None of the heralding probabilities  depends on the pump-phase offset, because our measurement modules are only sensitive to the photon number. The impact of pump-phase offset will appear when we calculate the fidelity of entanglement.

\subsection{Fidelity of DLCZ Entanglement Distribution}

The DLCZ fidelities of entanglement realized with PNRD and NRPD systems are
\begin{eqnarray}
\lefteqn{F_j  \equiv  \langle \psi _j |\rho _{{\rm pm}_j}|\psi _j \rangle \nonumber }\\
& =&  \frac{1}{{P_j }}\int \!{\frac{{{\rm d}^2 \zeta _a^L }}{\pi }\int \!{\frac{{{\rm d}^2 \zeta _a^R }}{\pi }} \left( {1 - \left| {\zeta _a^L  + ( - 1)^j \zeta _a^R } \right|^2 /2} \right)} \nonumber \\[.1in]
&\,& \times \int \!{\frac{{{\rm d}^2 \zeta _{p1} }}{\pi }\int \!{\frac{{{\rm d}^2 \zeta _{p2} }}{\pi }} } \,
\chi _A^{\rho _{\rm out} } (\zeta _a^L ,\zeta _a^R ,\zeta _{p1} ,\zeta _{p2} )
\nonumber \\[.1in] 
&\,&\times \left( {1 - \left| {\zeta _{pj} } \right|^2 } \right), \label{(B11)}
\end{eqnarray}
for $j=1,2$ in the PNRD case,
and 
\begin{eqnarray}
F_j  &=&   \frac{1}{{P_j }}\int \!{\frac{{{\rm d}^2 \zeta _a^L }}{\pi }\int \!{\frac{{{\rm d}^2 \zeta _a^R }}{\pi }} \left( {1 - \left| {\zeta _a^L  + ( - 1)^j \zeta _a^R } \right|^2 /2} \right)} \nonumber \\
&\,& \times \int \!{\frac{{{\rm d}^2 \zeta _{p1} }}{\pi }\int \!{\frac{{{\rm d}^2 \zeta _{p2} }}{\pi }} } \,
\chi _A^{\rho _{\rm out} } (\zeta _a^L ,\zeta _a^R ,\zeta _{p1} ,\zeta _{p2} )
\nonumber \\ 
&\,& \times \left( {\pi \delta (\zeta _{pj} ) - 1} \right),  
\label{(B16)}
\end{eqnarray}
for $j=1,2$ in the NRPD case, 
where we have used
\begin{equation}
\label{(E17)} 
\langle \psi _j | D_N (S_L ,\zeta _a^L )D_N (S_R ,\zeta _a^R )|\psi _j \rangle  = 1 - \frac{\displaystyle \left| \zeta _a^L  + ( - 1)^j \zeta _a^R  \right|^2 }{\displaystyle 2}.
\end{equation}
Both Eqs.~(\ref{(B11)}) and (\ref{(B16)}) can be evaluated via moment analysis from the Gaussian nature of Eq.~(\ref{(C1)}), yielding 
\begin{eqnarray}
\label{(14)}
\lefteqn{F_j = [\eta_j (1 - p_{c_L})(1 - p_{c_R})/(4P_j)]} \nonumber \\[.05in]
& \,& \times (\eta_L p_{c_L} + \eta_R p_{c_R} + 2 \sqrt{\eta_L p_{c_L}\eta_R p_{c_R}}
\cos(\theta_L-\theta_R)), \nonumber \\ [.05in]
&& \quad \mbox{for $j =1,2$,}
\end{eqnarray}
where for each detection scheme we use its corresponding heralding probability $P_j$.  Note that $F_j P_j$ is identical for both PNRD and NRPD systems. This can be qualitatively justified as follows. $F_j$ is the conditional probability of a successful entanglement creation given that a heralding event has occurred. Hence, $F_j P_j$ is the joint probability of successfully loading the ensembles in state $|\psi_j \rangle$ and the occurrence of the $M_j$ event. This joint event occurs when one---and only one---of the ensembles undergoes a single Raman transition to produce exactly one photon, and this photon is detected by photodetector $D_j$.  Photon-number resolution is not required for detecting a single photon, therefore both PNRD and NRPD systems have the same likelihood of a loading success.  It follows that the success probability, $P_{\rm success}$, is the same for the PNRD and NRPD systems, so in the appendix we will only present a derivation of Eq.~(\ref{(14)}) for the PNRD case. 
 
The fidelity in Eq.~(\ref{(14)}) is independent of which detector has clicked, provided that the detectors have the same efficiency, viz. $\eta_1 = \eta_2$. In this case, we have $P_{\rm success} = F_E P_{\rm herald}$, where $F_E \equiv F_1 = F_2$. (Here, the subscript $E$ emphasizes that we are concerned with the fidelity of entanglement.)  This means that the lower heralding probability of the PNRD system, relative to that of its NRPD counterpart, is exactly compensated by its higher fidelity of entanglement.  

   \begin{figure}
   \begin{center}
   \includegraphics[width=2.46in]{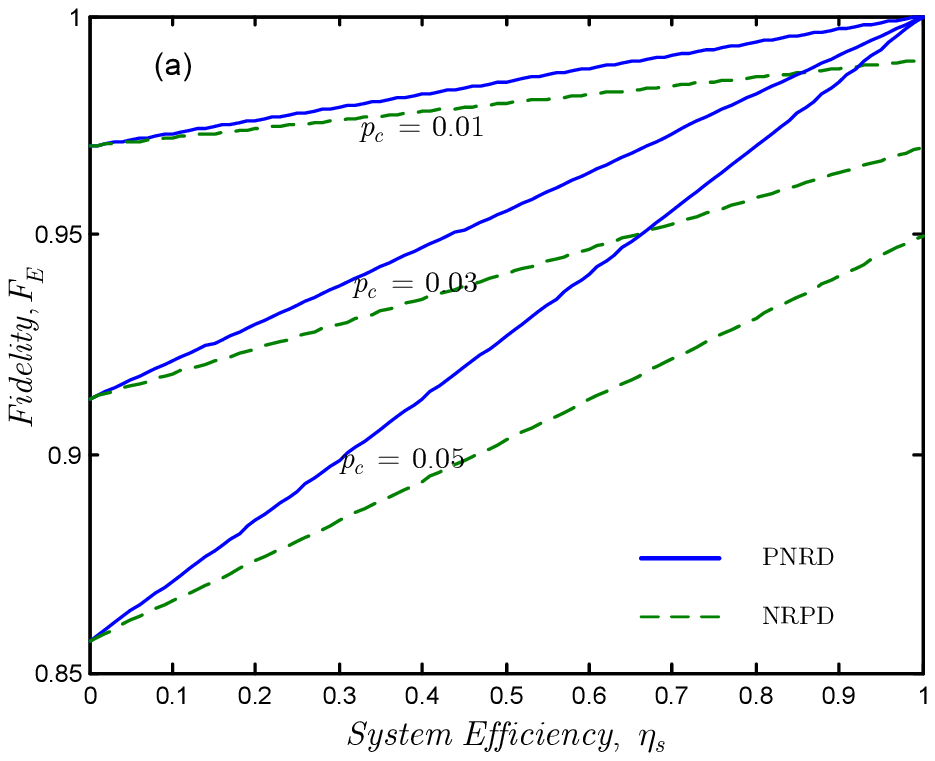}
   \end{center}
   \begin{center}
   \includegraphics[width=2.43in]{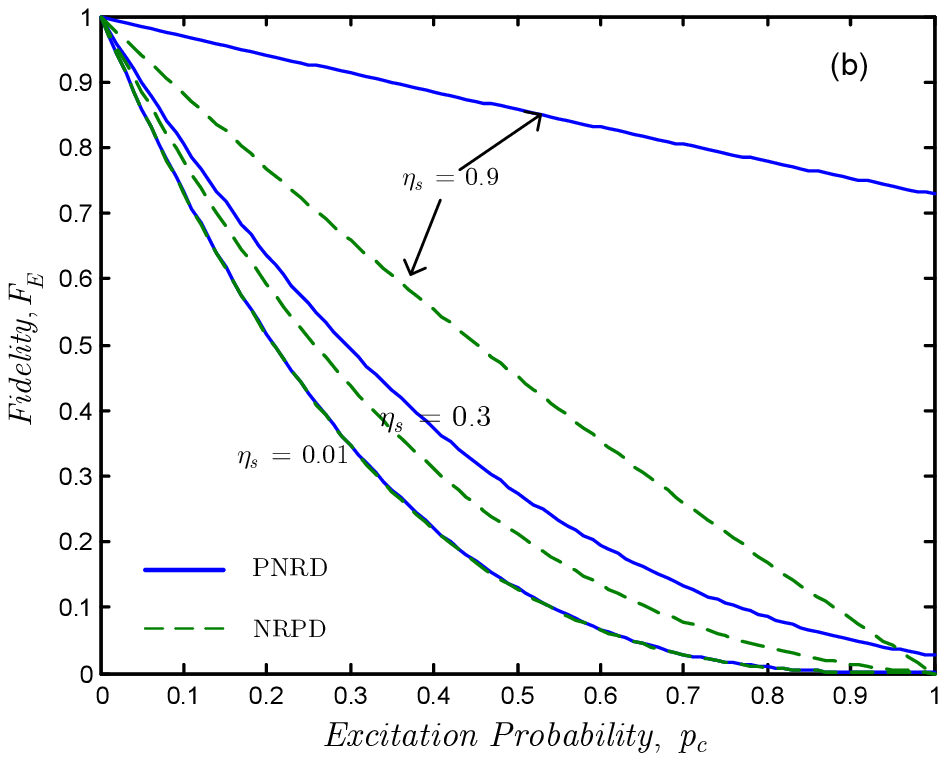}
   \end{center}
   \caption   { \label{F5} 
(Color online) Fidelity of entanglement, $F_E$, versus (a) system efficiency, $\eta_s$, and (b) excitation probability, $p_c$, for DLCZ entanglement distribution.  In both (a) and (b), we assume that the system setup is symmetric.}
   \end{figure} 

It is interesting to compare the behavior of the NRPD and PNRD fidelities of entanglement as we vary key system parameters.  For this purpose, it is easier to consider what happens in the symmetric case, when everything is identical for both ensembles and their corresponding Raman photons.  We then have
\begin{equation}
\label{PSsym}
P_{\rm success} = 2 \eta_s p_c (1-p_c)^2 , \quad\mbox{symmetric setup}
\end{equation}
and
\begin{equation}
F_{E,{\rm sym}} = \left\{\begin{array}{ll}
(\eta _s p_c  + 1 - p_c )^3 , &\mbox{PNRD}\\[.1in]
(1 - p_c )(\eta _s p_c  + 1 - p_c )^2, & \mbox{NRPD}.
\end{array}\right.
\label{FEfinal}
\end{equation}
The success probability of a symmetric setup, given by Eq.~({\ref{PSsym}), can also be obtained by the following simple argument.  A success occurs whenever one---and only one---of the ensembles produces a Raman photon and this photon is detected. In Eq.~({\ref{PSsym}), $P_{\rm success}$ is the product of $p_c (1-p_c)$ (the probability of one excitation) times $1-p_c$ (the probability of no excitations) times $\eta_s / 2$ (the survival probability for one photon) times 4 (the number of possibilities, all equiprobable, for emitting a single photon and getting a detector click).

From Eq.~(\ref{FEfinal}), we see that both the PNRD and NRPD $F_E$ expressions approach $(1 - p_c )^3  \simeq 1 - 3p_c $ for $p_c  \ll 1$ as the system efficiency $\eta _s $ approaches zero; this limit is in accord with preliminary results reported in the DLCZ paper \cite{qcomm4}.  In Fig.~\ref{F5}(a), we have plotted $F_E$ versus $\eta_s$ for the PNRD and NRPD systems.  From this figure we see that the PNRD system realizes perfect fidelity in the absence of loss ($\eta_s = 1$), whereas $F_E = 1-p_c$ for lossless operation of the NRPD system.  Figure~\ref{F5}(b) shows that the NRPD system is more sensitive to excitation probability ($p_c $) variations than is the PNRD system.  For $p_c \ll 1$, both systems approach perfect fidelity, but significant fidelity degradations occur for larger values of $p_c $. Indeed, from Eq.~(\ref{FEfinal}), we find that the NRPD system has zero fidelity at $p_c = 1$, whereas the PNRD system achieves  $F_E = \eta _s^3 $. Overall, in a practical operating regime in which $p_c \approx 0.01$ and $\eta_s \approx 0.01-0.1$ prevail, the PNRD and NRPD systems have very similar entanglement-distribution performance. This is important because NRPD technology is more advanced than PNRD technology.

\subsection{Asymmetric Setup and State Preparation}
\label{Secasym}

DLCZ entanglement distribution in an asymmetric configuration can be looked at in two different, but interrelated, ways. The first, which is the approach we have taken in deriving Eq. (\ref{(14)}), is to quantify the asymmetry-induced fidelity loss with respect to the maximally-entangled (singlet or triplet) states.  Deviations from complete symmetry, however, will make one path more probable than the other, and/or introduce relative phase terms.  Hence, the pure state for the two ensembles that is the best fit to their post-heralding joint density operator is, in general, a partially-entangled state of the form $d_L |1\rangle_L |0\rangle_R + d_R |0\rangle_L |1\rangle_R$, where $d_L$ and $d_R$ are functions of system parameters. This leads us to the second point of view, i.e., finding the most-likely (maximum-fideltiy) pure state for the asymmetric setup. The answer to this question provides us with a prescription for preparing the two ensembles in an arbitrary partially-entangled state.  In the appendix we show that the fidelity-maximizing state is 
\begin{eqnarray}
\label{optstate}
|\psi_j \rangle_{\rm opt} &=& \frac {\sqrt{\eta_L p_{c_L}}} {\sqrt {\eta_L p_{c_L} + \eta_R p_{c_R}}} |1 \rangle_L |0 \rangle_R + (-1)^j e^{i(\theta_R - \theta_L)}\nonumber \\ 
&\times& \frac {\sqrt{\eta_R p_{c_R}}} {\sqrt {\eta_L p_{c_L} + \eta_R p_{c_R}}} |0 \rangle_L |1 \rangle_R, \;\mbox{for $j=1,2$,} 
\end{eqnarray}
and the fidelity maximum that it achieves is 
\begin{eqnarray}
\label{FEopt}
F_{j,{\rm opt}} &\equiv & {}_{\rm opt} \langle \psi _j |\rho _{{\rm pm}_j}|\psi _j \rangle_{\rm opt} \nonumber \\[.05in]
& = & \frac{\eta_j (1-p_{c_L})(1-p_{c_R})  {\left( \eta_L p_{c_L}  + \eta_R p_{c_R} \right) }}{2 P_j }, 
\end{eqnarray}
for $j=1,2$. This is an intuitive result. The joint probability that $D_j$ clicks and that this click heralds successful loading of the state $|\psi_j\rangle_{\rm opt}$ is $ P_jF_{j,{\rm opt}}$,  which is given by the probability, $p_{c_{L/R}} (1-p_{c_L})(1-p_{c_R})$, of having exactly one excitation in only the left/right ensemble  times the probability, $\eta_j \eta_{L/R} /2$, that the associated Raman photon is detected by $D_j$.

A similar argument holds for the optimum state in Eq.~(\ref{optstate}). Here, the ratio between the probability of being in state $|1 \rangle _L |0 \rangle _R $ rather than in state $|0 \rangle _L |1 \rangle _R $ is $\eta_L p_{c_L} / (\eta_R p_{c_R})$, as expected. This ratio does not depend on the detector efficiencies, because the 50/50 beam splitter gives Raman photons an equal chance to be directed to $D_1$ or $D_2$.  On the other hand, the coherence between states $|1 \rangle _L |0 \rangle _R $ and $|0 \rangle _L |1 \rangle _R $ is impacted by the pump-phase offset difference between the two ensembles, as accounted for by the term $\exp[i(\theta_R - \theta_L)]$.

Figure~\ref{hood}(a) plots the optimum fidelity versus $\eta_L$ and $\eta_R$ for the PNRD case. Here, we assume all other parameters are the same for both ensembles. We see that the optimum fidelity degrades in response to decreasing either $\eta_L$ or $\eta_R$. Path loss affects fidelity in a PNRD system when multiple-excitation events are possible because loss allows multiple-photons events to masquerade as single-photon events, which can erroneously herald for success.  Therefore, when there is no path loss in a PNRD system its fidelity is unity. 

\begin{figure}
\begin{center}
 \includegraphics[width=2.5in]{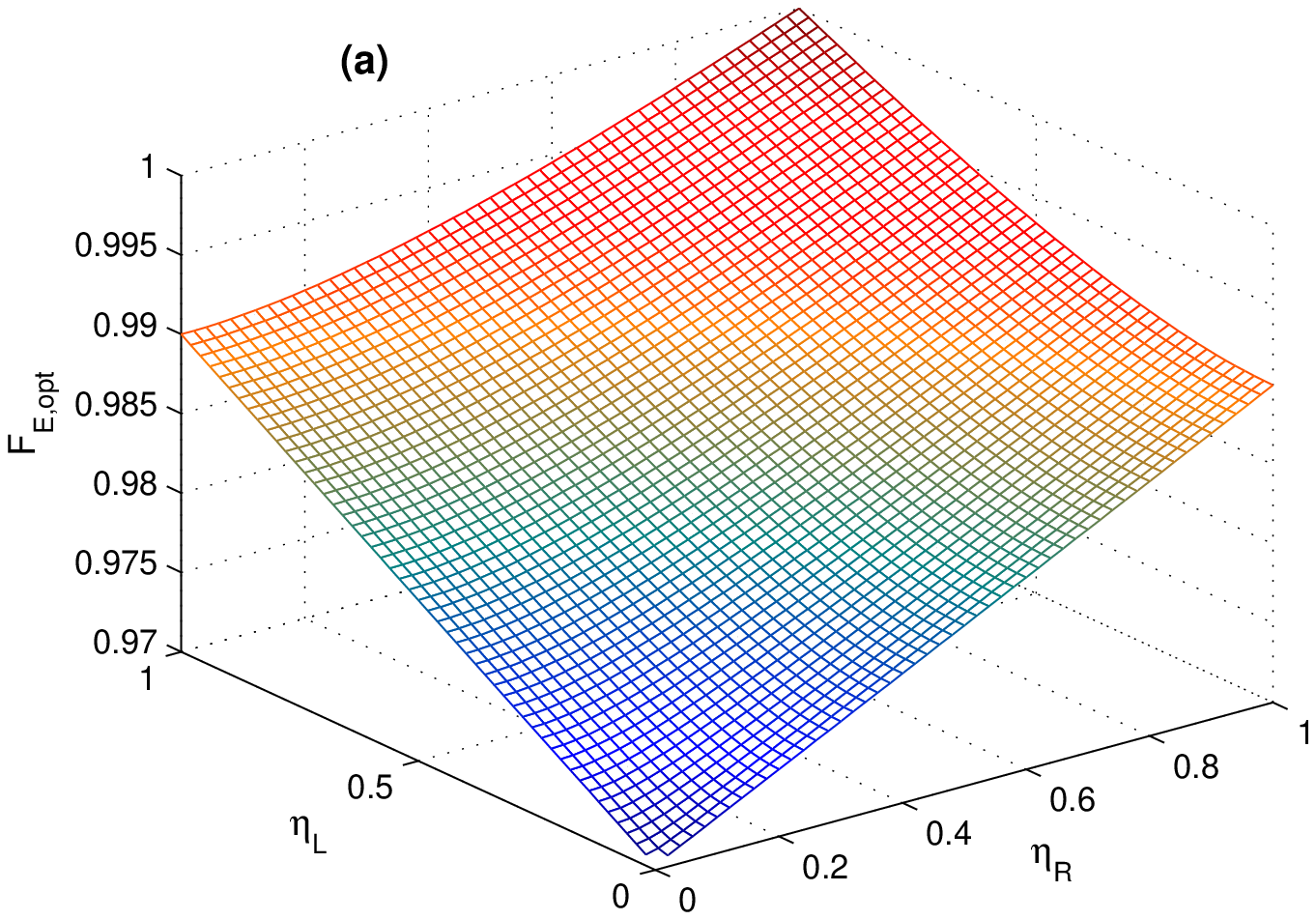}
 \end{center}
 \begin{center}
\includegraphics[width=2.5in]{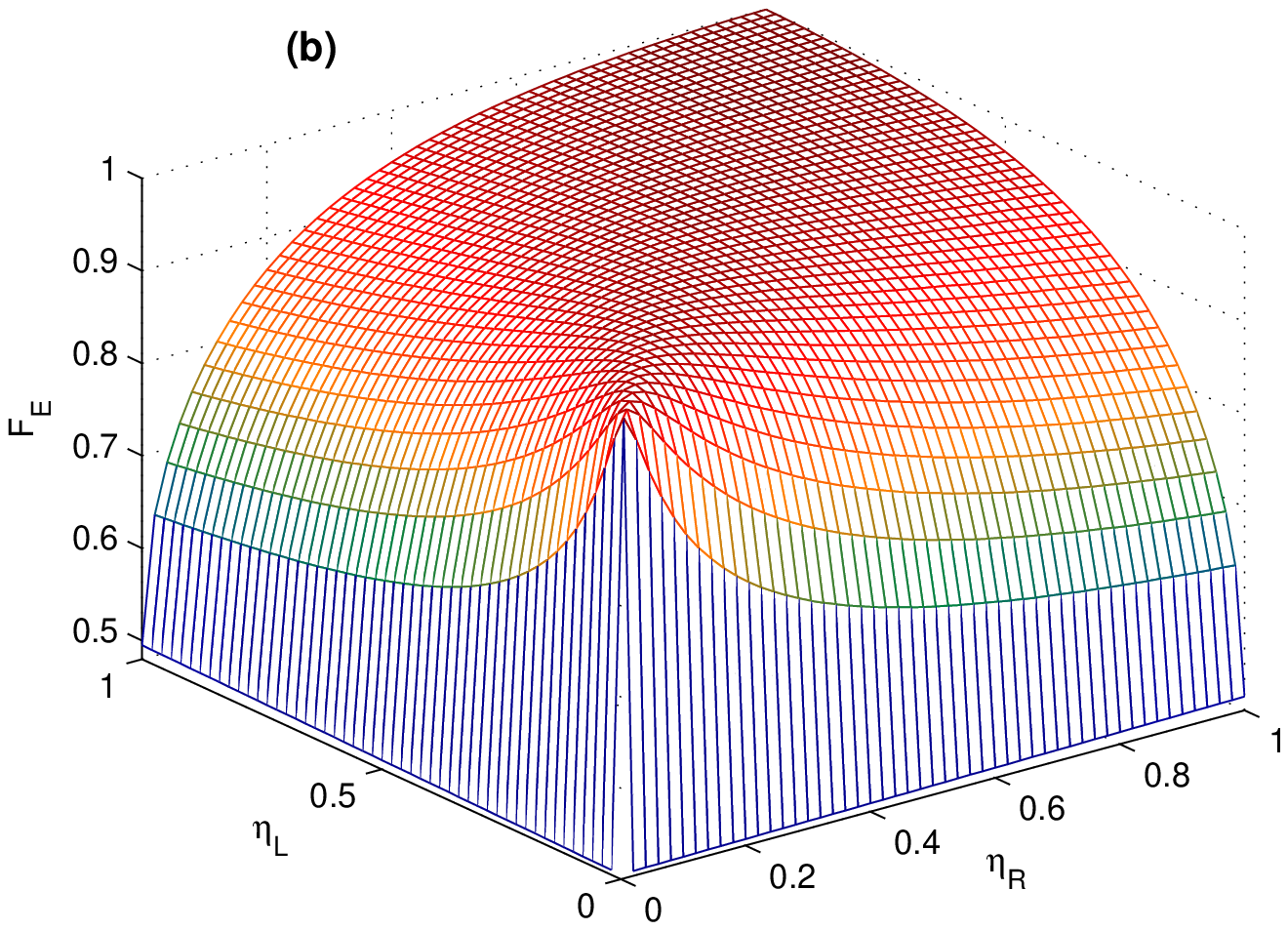}
 \end{center}
 \begin{center}
\includegraphics[width=2.5in]{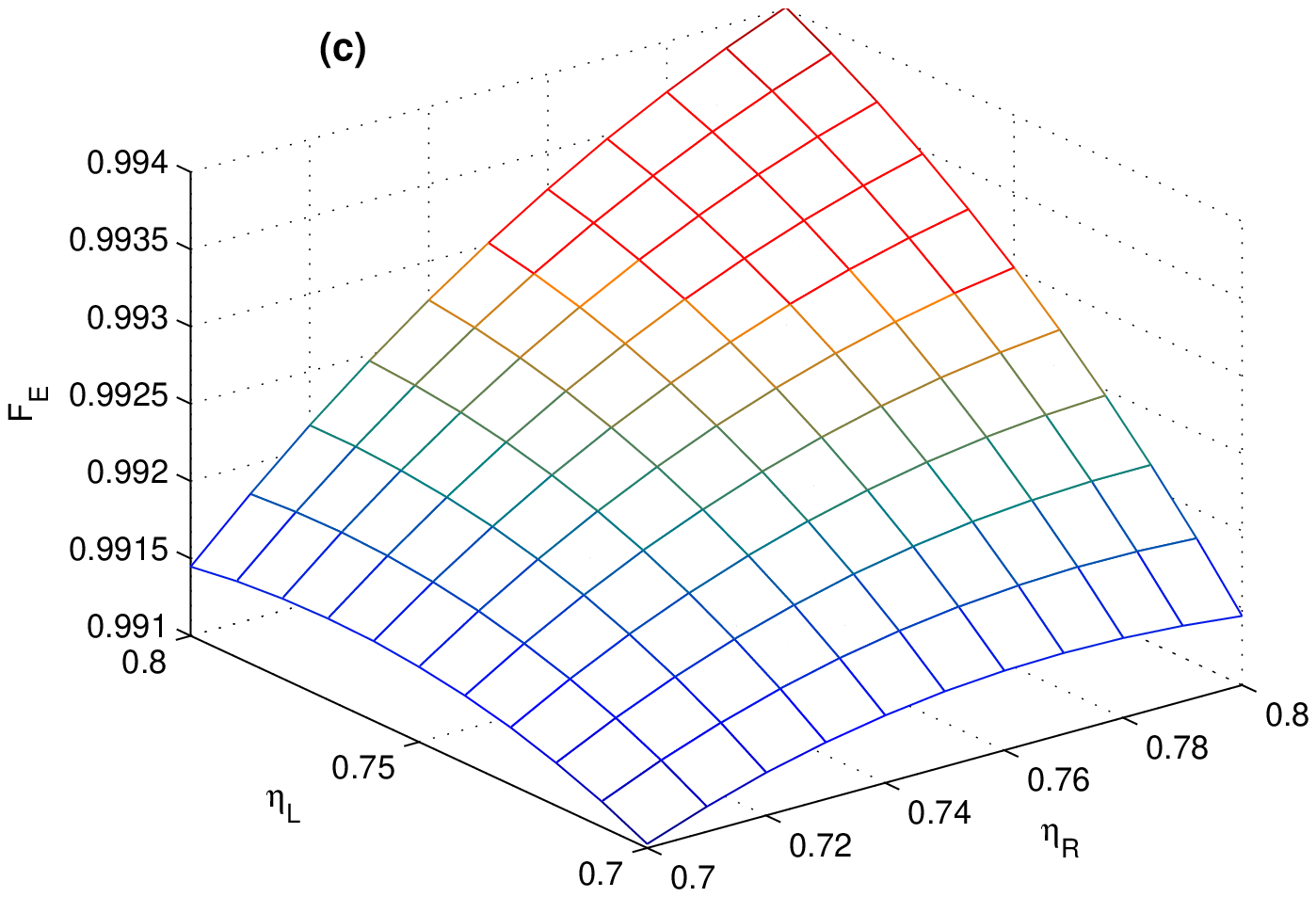}
 \end{center}
   \caption   { \label{hood} 
(Color online) (a) Optimum fidelity of entanglement for a DLCZ system with asymmetric path loss. In this case, the optimum (fidelity-maximizing) state is partially entangled. (b) and (c) Fidelity of entanglement (for a singlet/triplet state) versus left-path and right-path efficiencies, for DLCZ entanglement distribution. In all plots the only system asymmetry is $\eta_L\neq \eta_R$, and $p_c =0.01$, $\eta_1 = \eta_2 =1$ are assumed.}
   \end{figure} 

The degradation in the fidelity of entanglement arising from path-loss asymmetry, from Eq.~(\ref{(14)}), is shown in Fig.~\ref{hood}(b) to be increasingly severe as either $\eta_L$ or $\eta_R$ tends to zero. In this extreme case, we have almost complete which-path information on a photon detection; hence, noting that $F_j = |\langle \psi_j | \psi_j \rangle _ {\rm opt} |^ 2 F_{j, {\rm opt}}$, the fidelity becomes approximately $1/2$.  The asymptote is slightly less than 1/2, owing to multiple-excitation errors. Greater tolerance for path-loss asymmetry occurs at high values of $\eta_L$ and $\eta_R$, with asymmetry sometimes increasing the fidelity. Figure~\ref{hood}(c) shows this effect in the vicinity of $\eta_L = \eta_R = 0.7$: for $\eta_R = 0.7$ the peak fidelity occurs at $\eta_L \approx 0.78$. This is due to the fact that the projection $|\langle \psi_j | \psi_j \rangle _ {\rm opt} |$ is still very close to one for $\{\eta_L = 0.78,\eta_R = 0.7 \}$, and that the value of $F_{j, {\rm opt}}$ evaluated at $\{\eta_L = 0.78,\eta_R = 0.7 \}$ is higher than its value for $\eta_L = \eta_R = 0.7 $. On the other hand, fidelity always decreases if we degrade the system efficiency in either path.  
   
Now let us examine the effect of pump-phase asymmetry in the absence of any other sources of asymmetry.  Equation~(\ref{(14)}) assumes that $\theta_L$ and $\theta_R$ are deterministic phase shifts.  Although systematic (deterministic) phase shifts may be present in a real system, it is more important to study the effects of random phase errors.  Presuming $\theta_L$ and $\theta_R$ to be independent, identically distributed, zero-mean, Gaussian random variables with common variance $\sigma^2_{\theta}$, we obtain
\begin{equation}
F_E = F_{E,{\rm sym}}[1+ \exp(-\sigma^2_\theta)]/2 \, ,
\label{FEtheta}
\end{equation}
by averaging Eq.~(\ref{(14)}) over these pump-phase statistics.  It follows that $\sigma_\theta^2 \ll 1$ is a necessary condition for achieving high fidelity of entanglement in the DLCZ protocol.

\section{MIT/NU versus DLCZ entanglement distribution} 
\label{Seccomp}
The MIT/NU architecture is a singlet-based system for qubit teleportation that uses a novel ultrabright source of polarization-entangled photon pairs \cite{ultrabright}, and trapped rubidium atom quantum memories \cite{trap} whose loading can be nondestructively verified \cite{qcomm2, trap}. Figure~\ref{FM1}(a) shows a schematic of this system:  QM$_1 $ and QM$_2$ are trapped rubidium atom quantum memories, each $L_0 $\,km away---in opposite directions---from a dual optical parametric amplifier (OPA) source. As the overall structure of this architecture and its preliminary performance analysis have been described in considerable detail elsewhere \cite{qcomm2,Brent,exp4}, we shall provide only a brief description sufficient to enable comparison with the DLCZ scheme.

   \begin{figure}[h]
   \begin{center}
   \includegraphics[width=3.25in]{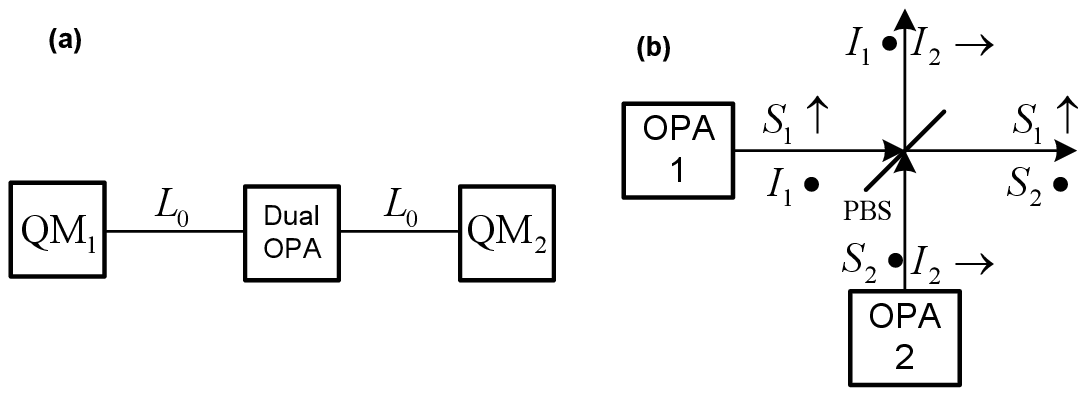}
   \end{center}
\begin{center}
\includegraphics[width=1.5in]{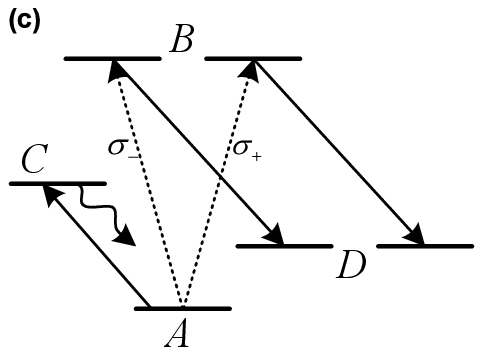}
\end{center}
\vspace*{-.2in}
   \caption
   { \label{FM1} 
(a) MIT/NU architecture for long-distance quantum communications consisting of a dual-OPA source that produces polarization-entangled photons, and two quantum memories, QM$_1$ and QM$_2$, separated by $2L_0$\,km. (b) Dual-OPA source of polarization-entangled photons.  OPAs~1 and 2 are coherently-pumped, continuous-wave, type-II phase matched, doubly-resonant amplifiers operated at frequency degeneracy whose orthogonally-polarized signal ($\{S_k\}$) and idler ($\{I_k\})$ outputs are combined, as shown, on the polarizing beam splitter (PBS). (c) Notional schematic for the relevant hyperfine levels of ${}^{87}$Rb. Each quantum memory consists of a single trapped rubidium atom that can absorb arbitrarily-polarized photons, storing their coherence in the long-lived $D$ levels. A non-destructive load verification is effected by means of the $A$-to-$C$ cycling transition.}
   \end{figure}

Each optical parametric amplifier in the dual-OPA source is a continuous-wave, type-II phase matched, doubly-resonant amplifier operating at frequency degeneracy.  Its signal ($S$) and idler ($I$) outputs comprise a stream of orthogonally-polarized photon pairs that are in a joint Gaussian state similar to Eq.~(\ref{(12)}) \cite{qcomm2}.  By coherently pumping two of these OPAs, and combining their outputs  on a polarizing beam splitter as shown in Fig.~\ref{FM1}(b), we obtain signal and idler beams that are polarization entangled \cite{ultrabright}.  These beams are routed down separate optical fibers to the trapped-atom quantum memories.  

A schematic of the relevant hyperfine levels of ${}^{87}$Rb is shown in Fig.~\ref{FM1}(c).  The memory atoms are initially in the ground state $A$.  From this state they can absorb a photon in an arbitrary polarization transferring that photon's coherence to the $B$ levels.  By means of a Raman transition, this coherence is shelved in the long-lived $D$ levels for subsequent use.  However, because propagation and fixed losses may destroy photons before they can be stored, and because both memories must be loaded with a singlet state prior to performing qubit teleportation, the MIT/NU architecture employs a clocked loading protocol in which the absence of fluorescence on the $A$-to-$C$ cycling transition provides a non-destructive indication that a memory atom has absorbed a photon.  If no fluorescence is seen from either the QM$_1$ or QM$_2$ atoms in a particular loading interval, then both memories have stored photon coherences and so are ready for the rest of the teleportation protocol, i.e., Bell-state measurements, classical communication of the results, and single-qubit rotations \cite{trap}.

A variety of error sources associated with the MIT/NU scheme have been identified and their effects analyzed \cite{Brent}.  Some are due to imperfections in the dual-OPA source, e.g. pump-power imbalance or pump-phase offsets between the two OPAs.  Others are due to the time-division multiplexed scheme---omitted from our brief description of the MIT/NU architecture---needed to compensate for the slowly-varying birefringence encountered in fiber propagation.  The most fundamental error source, however, is the same one we analyzed for the DLCZ protocol:  the emission of more than one pair of polarization-entangled photons, in conjunction with propagation and fixed losses, may lead to loading events (both memory atoms have absorbed photons) that do \em not\/\rm\ leave the memories in the desired singlet state.  This error mechanism is the primary one we shall consider here, although pump-phase offsets will also be included.

For a single trial of the MIT/NU loading protocol, let $P_{\rm herald}$ denote the probability that both memories are loaded, and let  $P_{\rm success}$ denote the probability that these memories have loaded the desired singlet state.  These probabilities are the MIT/NU counterparts to the heralding and success probabilities that we derived in Sec.~\ref{DLCZEntg} for DLCZ entanglement distribution.  Thus, for the MIT/NU architecture we have that $F_E = P_{\rm success}/P_{\rm herald}$ is its fidelity of entanglement.  From the work of Yen and Shapiro \cite{Brent}, we obtain
\begin{equation}
\label{PSMIT}
P_{\rm success}  =  \frac{{N^2  + \tilde n^2 [1 + \cos(\theta_1 -\theta_2 )] }}{{[(1 + \bar n)^2  - \tilde n^2 ]^4 }},
\end{equation}
and
\begin{equation}
\label{FidMIT}
F_E = \frac{{N^2  + \tilde n^2 [1 + \cos(\theta_1 -\theta_2 )]}}{{4N^2  + 2\tilde n^2 }},
\end{equation}
where 
\begin{equation}
N = \bar n(1 + \bar n) - \tilde n^2 ,\; 
\bar n =  I_ -   - I_ +, \;\mbox{ and }\;
\tilde n = I_ -   + I_ +, 
\end{equation}
with
\begin{equation}
I_ \pm   = \frac{{\eta _f \gamma \gamma _c }}{{\Gamma \Gamma _c }}\frac{{\left| G \right|}}{{(1 \pm \left| G \right|)(1 \pm \left| G \right| + \Gamma _c /\Gamma )}}.
\end{equation}
In these expressions:  the $\{\theta_j\}$ are the pump-phase offsets for the two OPAs; $\left| G \right|^2 $ is the normalized OPA pump gain ($|G|^2 = 1$ at oscillation threshold); $\Gamma $ and $\gamma $ are the OPA cavity's linewidth and its output coupling rate; $\Gamma _c $ and $\gamma _c $ are the memory cavity's linewidth and its input coupling rate; and $\eta _f $ is the transmissivity of the $L_0$-km-long source-to-memory fiber propagation path.  

   \begin{figure}
   \begin{center}
   \includegraphics[width=3in]{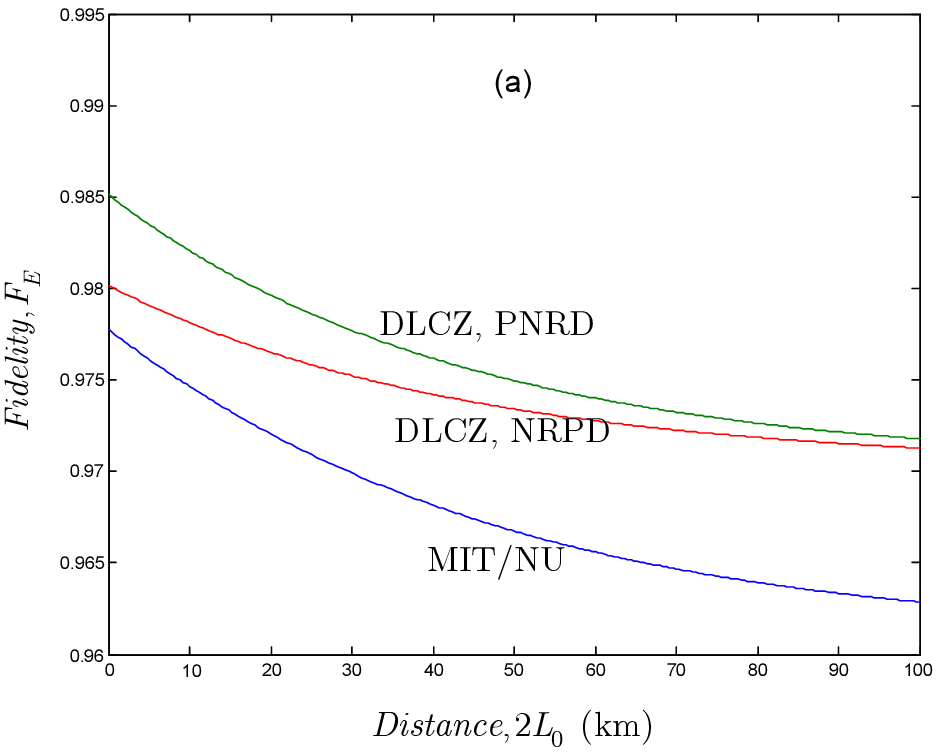}
   \end{center}
   \begin{center}
   \includegraphics[width=3in]{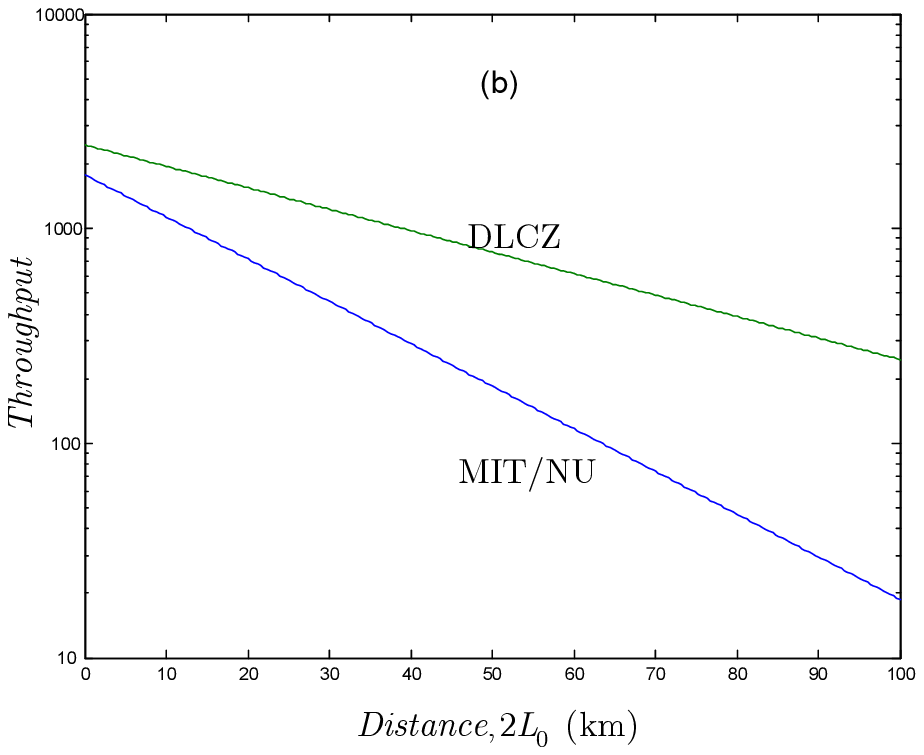}
   \end{center}
   \vspace*{-.2in}
   \caption
   { \label{FM2} 
(Color online) Performance comparison of the MIT/NU and DLCZ  entanglement-distribution architectures. (a) Fidelity of entanglement versus total distance between quantum memories in km. (b) Throughput (entangled pairs/sec) versus total distance between quantum memories in km.  The parameter values assumed in these plots are given in the text.}
   \end{figure} 

Using Eqs.~({\ref{PSsym}) and (\ref{FEfinal}) for the DLCZ protocol, and Eqs.~(\ref{PSMIT}) and (\ref{FidMIT}) for the MIT/NU architecture,  let us compare the behaviors of the fidelities and throughputs of entanglement for these two systems.  The latter, defined to be $RP_{\rm success}$, where $R$ is the rate at which either protocol is run, presumes that there are arrays of atomic ensembles (for DLCZ entanglement distribution) or trapped-atom quantum memories (for the MIT/NU architecture) that are loaded in succession.   In Fig.~\ref{FM2}(a) we have plotted the fidelities of entanglement versus the total distance $2L_0$ (in km) between the two atomic ensembles (DLCZ) or the two quantum memories (MIT/NU), and in Fig.~\ref{FM2}(b) we have plotted the associated throughputs.  The DLCZ curves assume the following parameter values:  zero pump-phase offsets; $p_c = 0.01$ excitation probability; $\eta_L = \eta_R$ corresponding to 0.2\,dB/km fiber loss; $\eta_1 = \eta_2 = 0.5$, and $R =  500$\,kHz.  The MIT/NU curves assume:  zero pump-phase offsets; $|G|^2 = 0.01$; $\eta_f$ corresponding to 0.2\,dB/km fiber loss; $\gamma\gamma_c/\Gamma\Gamma_c = 10^{-0.5}$ (5\,dB fixed loss per source-to-memory path);  $\Gamma_c/\Gamma = 0.5$; and $R = 500$\,kHz.  [Note that $p_c = 0.01$ for the DLCZ protocol is an equivalent source rate to $|G|^2 = 0.01$ for the MIT/NU architecture.]  

Figure~\ref{FM2}(a) shows that the DLCZ protocol has a slight advantage in fidelity of entanglement as compared to the MIT/NU architecture.  This advantage, however, may well disappear due to random pump-phase offsets.  In particular, if we let $\theta_1$ and $\theta_2$, in the MIT/NU architecture, be independent, identically-distributed, zero-mean Gaussian random variables with common variance $\sigma_\theta^2$, then averaged over this randomness the fidelity of entanglement from Eq.~(\ref{FidMIT}) reduces to
\begin{equation}
F_E = \frac{{N^2  + \tilde n^2 [1 + \exp(-\sigma^2_\theta )]}}{{4N^2  + 2\tilde n^2 }},
\end{equation}
which should be compared with Eq.~(\ref{FEtheta}).  Superficially, it would seem that both the DLCZ and MIT/NU systems suffer similar pump-phase offset degradations.  However, the MIT/NU architecture needs to stabilize the pump phases for two co-located OPAs, whereas the DLCZ protocol must stabilize the pump phases at a pair of atomic ensembles that are separated by a long distance ($2L_0$).  The latter task will surely be far more difficult than the former.

Figure~\ref{FM2}(b) shows that the DLCZ protocol has better throughput-versus-distance scaling than does the MIT/NU architecture.  This behavior has a simple physical explanation.  The DLCZ protocol relies on one Raman photon successfully traversing a distance $L_0$ and being detected, whereas the MIT/NU architecture requires two photons---a signal photon \em and\/\rm\ an idler photon---to successfully traverse a distance $L_0$ and be stored.  It should be noted, however, that all applications of the DLCZ scheme require two pairs of entangled ensembles \cite{qcomm4}. That reduces the effective throughput of the system by a multiplicative factor of 1/2.
\section{Quantum Communication with Atomic Ensembles} 
\label{Sectel}

In this section, we study some quantum communication applications of entangled atomic ensembles, as proposed in \cite{qcomm4}. Given that the prescription described in Sec.~\ref{DLCZEntg} provides high fidelity of entanglement ensembles, we will assume that ideal, maximum entanglement has been established between any two ensembles of interest in the quantum communication analyses that follow.  We could, instead, start our quantum communication studies from the joint density operator for the post-heralded state---found by accounting for multiple-excitation events by means of Gaussian-state analysis---for each pair of ensembles that has undergone DLCZ entanglement distribution.  It can be shown, however, that such an approach is unnecessary so long as the overall quantum communication performance is dominated by other parameters, such as loss in the measurement modules.  

\subsection{Quantum Repeaters and Entanglement Swapping}
\label{Secrep}

Truly long-distance quantum communication, e.g. for transcontinental applications, will require quantum repeaters to enable entanglement distribution over such extraordinary distances. This can be done by performing entanglement swapping  \cite{swap} on  two pairs of entangled ensembles in the cascade configuration shown in Fig.~\ref{repeater}(a). Here, ensembles $L_1$ and $R_1$ are entangled and $L$~km away from each other, as are $L_2$ and $R_2$, with $R_1$ and $L_2$ being co-located. Entanglement swapping can be done by performing a Bell-state measurement (BSM) on ensembles $R_1$ and $L_2$. This measurement entangles the $L_1$ and $R_2$ ensembles---separated by $2L$\,km---in a Bell state that is determined by the result of the BSM.

   \begin{figure}
   \begin{center}
      \includegraphics[width=3.25in]{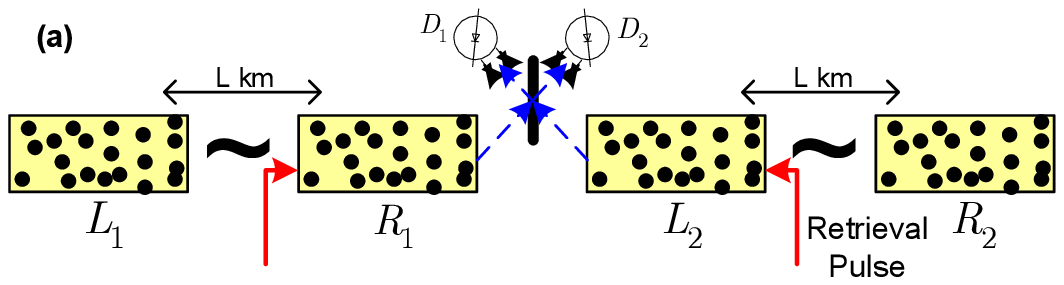}
      \end{center}
      \begin{center}
      \includegraphics[width=1in]{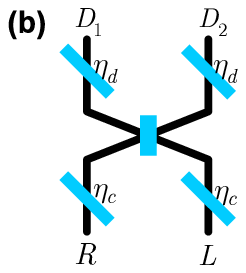}
      \end{center}
      \vspace*{-.2in}
   \caption
   { \label{repeater} 
(Color online) (a) DLCZ quantum repeater protocol. $\{L_1,R_1\}$ and $\{L_2,R_2\}$ are singlet states. By pumping $R_1$ and $L_2$ with strong retrieval pulses, we interfere any resulting anti-Stokes photons at a 50/50 beam splitter. Observing one---and only one---photon at one of detectors heralds protocol success, viz. $L_1$ and $R_2$ are now entangled. (b) Notional model for the measurement modules in (a):  beam splitters with vacuum-state quantum noise injected through their free input ports account for all loss and inefficiency effects; the single-photon detectors are assumed to have unity quantum efficiencies.}
   \end{figure} 

To perform a BSM on two atomic ensembles, we use detection of the anti-Stokes photons that can be produced by pumping the $|s\rangle\rightarrow |e\rangle$ transitions in the $\{ R_1, L_2 \}$ ensembles.  With strong retrieval pulses we can guarantee the emission of anti-Stokes ($|e\rangle\rightarrow |g\rangle$ transition) photons from every ensemble that was in its symmetric collective atomic state. Because these photons will be emitted in well-defined spatial modes, they can be routed to a 50/50 beam splitter---as shown in Fig.~\ref{repeater}(a)---which is followed by two single-photon detectors (either NRPDs or PNRDs).  Full BSM is not possible using only linear optics \cite{noBSM}, so the Fig.~\ref{repeater}(a) measurement scheme can only provide a partial BSM determining only two---out of four---Bell states.  Observation of a single click on one, and only one, of the detectors $\{D_1,D_2\}$, heralds completion of the DLCZ quantum-repeater protocol.  It is therefore a conditional protocol, whose fidelity and probability of success will be derived in this section.

Without loss of generallity, we shall assume that $\{L_1,R_1\}$ and $\{L_2,R_2\}$ have been placed in singlet states, and focus our attention on the losses and detector inefficiencies in the measurement module shown in Fig.~\ref{repeater}(a). As we did in our treatment of DLCZ entanglement distribution, we shall model the losses and detector inefficiencies by beam splitters, of transmissivities $\eta_c$ and $\eta_d$, which inject vacuum-state quantum noise through their free input ports, and take the detectors to have unity quantum efficiencies, see Fig.~\ref{repeater}(b).  The initial state of all four ensembles is thus
\begin{eqnarray}
|\psi _ {in} \rangle &=& (| 1 \rangle _ {L_1} |0 \rangle _ {R_1} - | 0 \rangle _ {L_1} |1 \rangle _ {R_1} )  
\nonumber \\ 
&\otimes& (| 1 \rangle _ {L_2} |0 \rangle _ {R_2} - | 0 \rangle _ {L_2} |1 \rangle _ {R_2} ) / 2 .
\end{eqnarray}
The above state consists of four orthogonal-state terms, each producing an orthogonal state after passing through the linear module of Fig.~\ref{repeater}(b). To find the heralding and success probabilities of the repeater it therefore suffices to find the corresponding figures of merit for each of these terms. Then, because of symmetry in the measurement module, the repeater fidelity, $F_R$, is just the ratio $P_{\rm success} / P_{\rm herald}$. We will use $P_{ij}$ to denote the heralding probability---i.e., having a click on either $D_1$ or $D_2$ but not both---that is due to state $|i \rangle _ {R_1} |j \rangle _ {L_2}$. Then, defining $\eta_m \equiv \eta_c \eta _d$ to be the measurement efficiency, we have
\begin{eqnarray}
\label{PHrep}
P_{\rm herald} & = & (P_{00} + P_{10} + P_{01} + P_{11})/4 \nonumber \\[.05in]
& = & \left\{\begin{array}{ll}
 \eta _m (2 - \eta_m ) / 2 , &\mbox{PNRD}\\[.1in]
 \eta _m (1 - \eta_m/2 ) / 2, & \mbox{NRPD}
\end{array}\right.
\end{eqnarray}
and
\begin{eqnarray}
\label{PSrep}
P_{\rm success} & = & (P_{01} + P_{10})/4 \nonumber \\
& = & \eta_m / 2,
\end{eqnarray}
where we used
\begin{subequations}
\label{Pij}
\begin{eqnarray}
& P_{00} = 0 , & \\
& P_{01} = P_{10} = \eta_m , & \\[.05in]
\label{P11}
& P_{11} = \left\{\begin{array}{ll}
2 \eta _m (1 - \eta_m ) , &\mbox{PNRD}\\[.1in]
2 \eta _m (1 - \eta_m/2 ), & \mbox{NRPD}.
\end{array}\right.
\end{eqnarray}
\end{subequations}

The preceding results show that the main source of error in the system is due to $P_{11}$, i.e., when we have two indistinguishable photons at the input of the 50/50 beam splitter. In this case, the $L_1$ and $R_2$ ensembles are in their ground states after the herald occurs, and thus the heralding event does \em not\/\rm\ imply a successful entanglement swap.  That such an erroneous heralding can occur is due to quantum interference. When a pair of indistinguishable photons enter a 50/50 beam splitter---one through each input port---they undergo quantum interference that makes both exit from the same output port \cite{HOM}. Now, if we are using NRPDs, these two photons may reach one of the detectors with probability $\eta_m^2$ and erroneously herald for success. Note that a PNRD system can identify this type of error.   However, if one---and only one---of the two photons is absorbed en route to the PNRDs, then they too can be fooled into heralding an entanglement swap when no such swap has occurred. This loss event occurs with probability $2(1-\eta_c) \eta_m + 2 \eta_c^2 \eta_d (1-\eta_d)$ for both NRPD and PNRD systems. The sum of these probabilities results in Eq.~(\ref{P11}). It follows that the maximum fidelity, achieved at $\eta_m = 1$, of a PNRD-based repeater is unity, whereas for an NRPD-based system it is only $2/3$. In general, from Eqs.~(\ref{PHrep}) and (\ref{PSrep}), we obtain
\begin{eqnarray}
F_R & = & P_{\rm success} / P_{\rm herald} \nonumber \\[.05in]
& = & \left\{\begin{array}{ll}
1 / (2 - \eta_m ) , &\mbox{PNRD}\\[.1in]
1 / (2 - \eta_m/2 ), & \mbox{NRPD}.
\end{array}\right.
\end{eqnarray}

\subsection{DLCZ Teleportation}

The DLCZ teleportation scheme is a conditional protocol for teleporting a qubit from one pair of atomic ensembles to another, see Fig.~\ref{Fig2}, \cite{qcomm4}.  It assumes that ensembles $\{L_1 ,R_1 \}$ and $\{L_2 ,R_2 \}$ have each been entangled in singlet states by means of the entanglement distribution protocol described in Sec.~\ref{DLCZEntg}---perhaps augmented by quantum repeaters to achieve even longer distances than can be realized with by entanglement distribution alone---where ensembles $\{L_1, L_2\}$ are co-located, as are ensembles $\{R_1,R_2\}$, with the latter pair being a distance $L$ away from the former.  The qubit to be teleported is the state
\begin{eqnarray}
|\psi_{\rm in}\rangle_{I_1I_2} &\equiv& 
d_0|1\rangle_{I_1}|0\rangle_{I_2} + d_1|0\rangle_{I_1}|1\rangle_{I_2},\nonumber \\[.05in]
&& \quad\mbox{where $|d_0|^2 + |d_1|^2 = 1$},
\end{eqnarray}
stored in two other ensembles, $\{I_1,I_2\}$,  which are co-located with $\{L_1,L_2\}$.  Such a state can be prepared by using the asymmetric setup as discussed in Sec.~\ref{Secasym}. Our objective is to make a measurement that  transfers the $\{d_0,d_1\}$ coherence to the remote ensembles $\{R_1,R_2\}$.  

To accomplish this teleportation, we need two simultaneous entanglement swaps: a BSM on $L_1$ and $I_1$, and a BSM on $L_2$ and $I_2$.  As depicted in Fig.~\ref{Fig2}, the required BSM is performed by the same measurement module used in the DLCZ quantum repeater.  Thus, DLCZ teleportation is conditional, hence it can only be used if $\{I_1,I_2\}$ can be restored to the state $|\psi_{\rm in }\rangle$ when the heralding event fails to occur.  In what follows we will sketch a derivation of the fidelity of DLCZ teleportation, 
\begin{eqnarray}
F_T &\equiv&
P_+ \,{}_{R_1R_2} \langle\psi_{\rm out}^+|\rho_{\rm out} ^ +|\psi_{\rm out}^+\rangle_{R_1R_2}  
\nonumber \\[.05in] &+& 
P_- \,{}_{R_1R_2} \langle\psi_{\rm out}^-|\rho_{\rm out} ^ -|\psi_{\rm out}^-\rangle_{R_1R_2}, 
\label{FEstart}
\end{eqnarray}
where $P_+$ is the probability of heralding on $\{D_1^L,D_2^L\}$ or $\{D_1^I,D_2^I\}$, $P_-$ is the probability of heralding on $\{D_1^L,D_2^I\}$ or $\{D_1^I,D_2^L\}$, 
\begin{equation}
|\psi_{\rm out}^{\pm}\rangle_{R_1R_2} \equiv 
d_0|1\rangle_{R_1}|0\rangle_{R_2} \pm d_1|0\rangle_{R_1}|1\rangle_{R_2},
\end{equation}
are the desired output states for the $\{R_1,R_2\}$ ensembles, and $\rho_{\rm out}^{\pm}$ are their actual output states, conditioned on there being a $P_{\pm}$ heralding event.

   \begin{figure}
   \begin{center}
   \begin{tabular}{c}
   \includegraphics[width=3.25in]{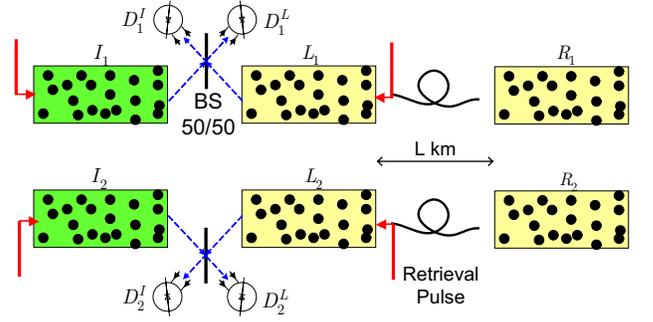}
   \end{tabular}
   \end{center}
   \caption
   { \label{Fig2} 
(Color online) DLCZ scheme for conditional teleportation. Two pairs of entangled atomic ensembles $\{L_1,R_1\}$ and $\{L_2,R_2\}$ are used to teleport the quantum state of ensembles $\{I_1,I_2\}$ to ensembles $\{R_1,R_2\}$. Strong retrieval pulses, which are near-resonant with the $|s\rangle\rightarrow|e\rangle$ transition, are used to pump ensembles $\{L_1,L_2,I_1,I_2\}$, recovering anti-Stokes photons from every ensemble that was in its symmetric collective atomic state.  Detection of a photon by one, and only one, of the single-photon detectors in each measurement module heralds completion of the protocol.  }
   \end{figure} 

The initial state of all six ensembles is 
\begin{eqnarray}
\label{(1)}
\left| {\psi _{\rm in} } \right\rangle &\equiv& \left( {d_0 \left| 1 \right\rangle _{I_1 } \left| 0 \right\rangle _{I_2 }  + d_1 \left| 0 \right\rangle _{I_1 } \left| 1 \right\rangle _{I_2 } } \right) \nonumber \\ 
&\otimes& \left( {\left| 0 \right\rangle _{L_1 } \left| 1 \right\rangle _{R_1 }  - \left| 1 \right\rangle _{L_1 } \left| 0 \right\rangle _{R_1 } } \right)/\sqrt 2 \nonumber \\
&\otimes& \left( {\left| 0 \right\rangle _{L_2 } \left| 1 \right\rangle _{R_2 }  - \left| 1 \right\rangle _{L_2 } \left| 0 \right\rangle _{R_2 } } \right)/\sqrt 2 {\rm\ .}
\end{eqnarray}
We can quickly home in on the output state $\rho_{\rm out}$ by multiplying out in Eq.~(\ref{(1)}), throwing away all terms that cannot lead to heralding, and then renormalizing.  The resulting ``short-form'' input state is
\begin{eqnarray}
\label{(2)}
\left| {\psi _{\rm in} } \right\rangle _{\rm short}  &=& - \frac{d_0}{\sqrt{2}}\left| 0 \right\rangle _{L_1 } \left| 1 \right\rangle _{I_1 } \left| 1 \right\rangle _{L_2 } \left| 0 \right\rangle _{I_2 }  \left| 1 \right\rangle _{R_1 } \left| 0 \right\rangle _{R_2 }  \nonumber \\ &-& \frac{d_1}{\sqrt{2}}\left| 1 \right\rangle _{L_1 } \left| 0 \right\rangle _{I_1 } \left| 0 \right\rangle _{L_2 } \left| 1 \right\rangle _{I_2 }  \left| 0 \right\rangle _{R_1 } \left| 1 \right\rangle _{R_2 }  \nonumber \\
 &+& \frac{d_0}{\sqrt{2}}\left| 1 \right\rangle _{L_1 } \left| 1 \right\rangle _{I_1 } \left| 1 \right\rangle _{L_2 } \left| 0 \right\rangle _{I_2 }  \left| 0 \right\rangle _{R_1 } \left| 0 \right\rangle _{R_2 }  \nonumber \\ &+& \frac{d_1}{\sqrt{2}}\left| 1 \right\rangle _{L_1 } \left| 0 \right\rangle _{I_1 } \left| 1 \right\rangle _{L_2 } \left| 1 \right\rangle _{I_2 } \left| 0 \right\rangle _{R_1 } \left| 0 \right\rangle _{R_2 } .
\end{eqnarray}

The success or failure of  DLCZ teleportation---given that a heralding event has occurred---can be understood by scrutinizing $|\psi_{\rm in}\rangle_{\rm short}$. A heralding event generated by the first two terms (the {\em good} terms) on the right-hand side of Eq.~(\ref{(2)}) yields the desired teleportation result, but a heralding event that is due to the last two terms (the {\em bad} terms) in this equation leaves the $\{R_1,R_2\}$ ensembles in their ground states.  Physically, it is easy to see what leads to this behavior.  Heralding that is due to the good terms results from exactly two photons being detected:  one from ensemble $L_1$ (or $I_1$) in the upper measurement module of Fig.~\ref{Fig2}(a), and one from ensemble $I_2$  (or $L_2$) in the lower measurement module in that figure.  The measurement-module beam splitters erase which-way information, and thus teleportation is completed.  Now, suppose that we have perfect measurement efficiency ($\eta_m \equiv \eta_c\eta_d = 1$) and consider what happens when the heralding is due to one of the bad terms.  In this case three photons enter the measurement modules:  either one each from $L_1$ and $I_1$ plus one from $L_2$, or one from $L_1$ and one each from $L_2$ and $I_2$.   In either case the $\{R_1,R_2\}$ ensembles are left in their ground states, hence the resulting $\rho_{\rm out}$ will be outside the Hilbert space spanned by $|\psi_{\rm out}^\pm\rangle_{R_1R_2}$.   So, whether or not the bad terms degrade DLCZ teleportation fidelity depends on whether the measurement modules can distinguish the good terms in Eq.~(\ref{(2)}) from the bad ones.

To evaluate the teleportation fidelity, we can use the heralding probabilities we obtained in Sec.~\ref{Secrep} along with the distinction we have drawn between good and bad terms to obtain
\begin{equation}
F_T = \left\{\begin{array}{ll}
1/(3-2\eta_m), & \mbox{PNRD}\\[.1in]
1/(3-\eta_m), & \mbox{NRPD},\end{array}\right.
\label{FT_DLCZ}
\end{equation}
where we used
\begin{eqnarray}
P_{\rm success} & = & P_{01} P_{10} /4 \nonumber \\
& = & \eta_m^2 / 4
\end{eqnarray}
and
\begin{eqnarray}
P_{\rm herald} & = & (P_{01} P_{10} + P_{11} P_{01} ) / 4 \nonumber \\
& = & \left\{\begin{array}{ll}
\eta_m^2 (3 - 2 \eta_m) / 4, & \mbox{PNRD}\\[.1in]
\eta_m^2 (3 - \eta_m ) / 4, & \mbox{NRPD.}\end{array}\right.
\label{Ph_DLCZ}
\end{eqnarray}
It follows that with perfect measurement efficiencies, the teleportation fidelity of the PNRD-based system is $F_T = 1$ and that of the NRPD-based system is $F_T = 1/2$.  
In Fig.~\ref{F4} we have plotted $F_T$ versus $\eta_m$ for the PNRD and NRPD cases.  The NRPD system never attains high fidelity because of its inability to suppress heralding from the bad terms in $|\psi_{\rm in}\rangle_{\rm short}$.  The PNRD does realize high teleportation fidelity, but only when its measurement efficiency is similarly high.  

   \begin{figure}
   \begin{center}
   \begin{tabular}{c}
   \includegraphics[width=3in]{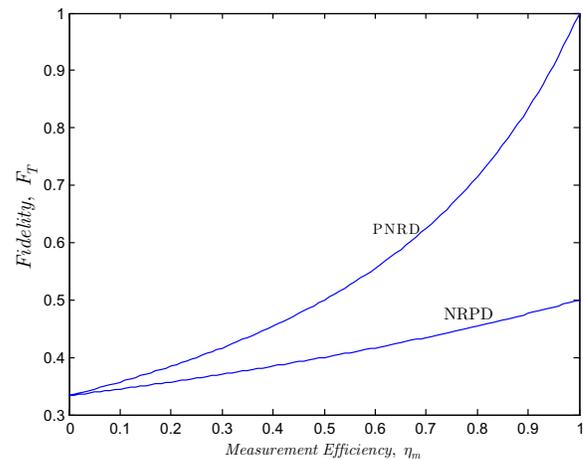}
   \end{tabular}
   \end{center}
   \caption
   { \label{F4} 
(Color online) Fidelity of DLCZ teleportation, $F_T$, versus measurement efficiency, $\eta_m$.}
   \end{figure}

DLCZ teleportation is rather different from MIT/NU teleportation.  The DLCZ approach is conditional, 
hence it can only be used if the $\{I_1,I_2\}$ ensembles in Fig.~\ref{Fig2}(a) can be restored to the state $|\psi_{\rm in }\rangle$ when the heralding event fails to occur.  The MIT/NU approach is unconditional, hence it is suitable for networking quantum computers.  On the other hand, the measurements required by the DLCZ scheme---high measurement-efficiency PNRD modules---seem significantly less challenging, given the current state of technology, than what is needed by the MIT/NU system, viz. Bell-state measurements on trapped atoms.  

\section{CONCLUSIONS}

We have compared the performance of DLCZ entanglement distribution, which is based on atomic ensembles, with that of the MIT/NU architecture, which relies on trapped-atom quantum memories.  We showed that the DLCZ protocol for entanglement distribution could achieve a better throughput-versus-distance behavior than the MIT/NU architecture, with both being capable of high fidelities of entanglement. In contrast, DLCZ quantum-repeater and teleportation protocols are conditional, and their performance depends critically on the availability of high-efficiency photon-number resolving photodetectors.  The MIT/NU teleportation system, on the other hand, is unconditional, but needs to realize Bell-state measurements within its trapped-atom quantum memories.  

\section*{ACKNOWLEDGMENTS} 
This work was supported in part by the Department of Defense Multidisciplinary University Research Initiative program under Army Research Office grant DAAD-19-00-1-0177, and 
by the HP-MIT Alliance.

\appendix*
\section{}
\label{App}

In this appendix, we derive the fidelity of entanglement for the DLCZ architecture. We assume photon-number resolving detectors (PNRDs) are being used in the detection setup, and we find the fidelity $F_{j,\rm{d}}$ of being in an arbitrary pure state $|\psi_{\rm d} \rangle = d_L |1 \rangle _ L |0 \rangle _ R + d_R |0 \rangle _ L |1 \rangle _ R $ after the occurrence of event $M_j$  as defined in Eqs.~(\ref{M1}) and (\ref{M2}). From Eq.~(\ref{(B8)}), and the fact that  $\langle \psi _{\rm d} | D_N (S_L ,\zeta _a^L )D_N (S_R ,\zeta _a^R )|\psi _{\rm d} \rangle  = 1 - | d_L^\ast \zeta _a^L  + d_R^\ast \zeta _a^R  |^2 $, we obtain
\begin{eqnarray}
F_{j,{\rm d}} &\equiv&  \langle \psi _{\rm d} |\rho _{{\rm pm}_j}|\psi _{\rm d} \rangle \nonumber \\
& =&  \frac{1}{{P_j }}\int \!{\frac{{{\rm d}^2 \zeta _a^L }}{\pi }\int \!{\frac{{{\rm d}^2 \zeta _a^R }}{\pi }} \left( {1 - \left| { d_L^\ast \zeta _a^L  + d_R^\ast \zeta _a^R } \right|^2} \right)} \nonumber \\[.1in]
&\,& \times \int \!{\frac{{{\rm d}^2 \zeta _{p1} }}{\pi }\int \!{\frac{{{\rm d}^2 \zeta _{p2} }}{\pi }} } \,
\chi _A^{\rho _{\rm out} } (\zeta _a^L ,\zeta _a^R ,\zeta _{p1} ,\zeta _{p2} ) 
\nonumber \\ &\,& \times \left( {1 - \left| {\zeta _{pj} } \right|^2 } \right),\quad\mbox{PNRD, $j= 1,2$},
\label{Fid}
\end{eqnarray}
where $P_j$ has been obtained in Eq.~(\ref{(B9)}). The key technique to evaluating the above integral lies in the Gaussian form of $\chi _A^{\rho _{\rm out} } (\zeta _a^L ,\zeta _a^R ,\zeta _{p1} ,\zeta _{p2} )$, as described in Eq.~(\ref{(C1)}). This function can be written in the following form
\begin{equation}
\chi _A^{\rho _{\rm out} } (\zeta _a^L ,\zeta _a^R ,\zeta _{p1} ,\zeta _{p2} ) = (2 \pi)^4 \sqrt{\det {\bf K}}\, G({\bm \zeta},{\bf K}),
\end{equation}
where
\begin{equation}
{\bm \zeta} = [\zeta _{ar}^L ,\zeta _{ai}^L ,\zeta _{pr}^ -  ,\zeta _{pi}^ -  ,\zeta _{pr}^ +  ,\zeta _{pi}^ +  ,\zeta _{ar}^R ,\zeta _{ai}^R ]^T ,
\end{equation}
\begin{equation}
G({\bm x},{\bf C}) = (2 \pi )^{-n/2} (\det {\bf C})^{-1/2} \exp { (- {\bm x}^T {\bf C}^{-1} {\bm x} / 2)},
\end{equation}
and ${\bm x} = [x_1,\ldots,x_n]^T$ is a real-valued column vector. The function $G({\bm x},{\bf C})$ represents the joint probability density function for $n$ zero-mean Gaussian random variables $X_1, \ldots,X_n$, with covariance matrix ${\bf C}$, evaluated at point $ {\bm x}$. The covariance matrix elements are ${\bf C}_{ij} = E_{\bm x}\{X_i X_j\}$, where $E_{\bm x}\{ \cdot\}$ denotes the statistical averaging over $X_1, \ldots,X_n$. With this new notation, the integral in Eq.~({\ref{Fid}) can be written as follows
\begin{eqnarray}
\label{Fid2}
F_{j,{\rm d}} &=& \frac {16 \sqrt {\det {\bf K}} } {\eta _1 \eta _2 P_j } E_{\bm \zeta}  \left\{ {1 - \left| {\zeta _{pj} } \right|^2  - \left| {d_L^\ast \zeta _a^L  + d_R^\ast \zeta _a^R } \right|^2 }\right.  \nonumber \\  &\,&
\left.{ + \left| {\zeta _{pj} } \right|^2 \left| {d_L^\ast \zeta _a^L  + d_R^\ast \zeta _a^R } \right|^2 }\right\},
\end{eqnarray}
where the factor $\eta_1 \eta_2$ is due to the change of variables from $\{ \zeta_{p1} , \zeta_{p2} \}$ to $\{ \zeta_p^+ , \zeta_p^- \}$ using Eq.~(\ref{zetapm}). The above moments can be written in terms of the elements of the covariance matrix ${\bf K}$. The latter can be found by inverting ${\bf K}^{-1}$, which can be easily obtained from Eq.~(\ref{(C1)}). The resulting symmetric matrix has been summarized in Table~\ref{Kmat}. It can be shown that $\sqrt{\det {\bf K}} = \eta_1 \eta_2 / (4 \alpha_L \alpha_R) $. Now, we can simplify Eq.~(\ref{Fid2}), by noting that
\begin{eqnarray}
\label{Fid3}
\lefteqn{E_{\bm \zeta} \left \{ | \zeta_{pj} |^2 \right \} = \frac{E_{\bm \zeta} \left \{ | \zeta_p^+ |^2 + | \zeta_p^- |^2 +2(-1)^j \Re \{\zeta_p^+ {\zeta_p^-}^\ast \} \right \}}{2 \eta_j}} \nonumber \\
& = & \frac{[{\bf K}_{55} + {\bf K}_{66} + {\bf K}_{33} + {\bf K}_{44} + 2(-1)^j ({\bf K}_{35} +{\bf K}_{46})]}{2 \eta_j} \nonumber \\
& = & 1 .
\end{eqnarray}
Also, by using the moment-factoring theorem for Gaussian variables, we obtain
\begin{eqnarray}
\lefteqn{E_{\bm \zeta}  \left\{ \left| {\zeta _{pj} } \right|^2 \left| {d_L^\ast \zeta _a^L  + d_R^\ast \zeta _a^R } \right|^2  \right \}}  \nonumber \\ &=&  \left | E_{\bm \zeta}  \left \{ \zeta _{pj} (d_L^\ast \zeta _a^L  + d_R^\ast \zeta _a^R ) \right \} \right| ^2 \nonumber \\ 
&\,&  + \left | E_{\bm \zeta}  \left\{ \zeta _{pj}^\ast (d_L^\ast \zeta _a^L  + d_R^\ast \zeta _a^R ) \right \} \right| ^ 2 \nonumber \\
&\,& + E_{\bm \zeta} \left \{ | \zeta_{pj} |^2 \right \}
E_{\bm \zeta} \left \{ \left| {d_L^\ast \zeta _a^L  + d_R^\ast \zeta _a^R } \right|^2  \right \} ,
\end{eqnarray}
in which
\begin{eqnarray}
\lefteqn{\hspace*{-.2in}E_{\bm \zeta}  \left \{ \zeta _{pj} (d_L^\ast \zeta _a^L  + d_R^\ast \zeta _a^R ) \right \} 
= }\nonumber \\ &&\hspace*{-.25in} \sqrt{\frac{\displaystyle \eta_j}{\displaystyle 2}} \left( (-1)^{j-1} \sqrt{ \eta_L p_{c_L} } d_L^\ast e^{i \theta_L}-  \sqrt{ \eta_R p_{c_R} } d_R^\ast e^{i \theta_R} \right)
\end{eqnarray}
and
\begin{equation}
\label{Fid4}
E_{\bm \zeta}  \left \{ \zeta _{pj}^\ast (d_L^\ast \zeta _a^L  + d_R^\ast \zeta _a^R ) \right \} = 0 .
\end{equation}
\begin{table*}
\caption{\label{Kmat} The elements of the covariance matrix ${\bf K}$.}
		\begin{ruledtabular}
		\begin{tabular} {c}
 ${\bf K}_{11} = {\bf K}_{22} =  (1-p_{c_L})/2 + \eta_L p_{c_L} (\eta_1 + \eta_2) / 4$ \\
 ${\bf K}_{24} = {\bf K}_{42} = - {\bf K}_{13} = - {\bf K}_{31} = (\eta_1+\eta_2) \sqrt{\eta_L p_{c_L} } \cos \theta_L / 4 $ \\
 ${\bf K}_{14} = {\bf K}_{23} = {\bf K}_{32} = {\bf K}_{41} = -(\eta_1+\eta_2) \sqrt{\eta_L p_{c_L} } \sin \theta_L / 4$ \\
$ {\bf K}_{15} = {\bf K}_{51} = - {\bf K}_{26} = - {\bf K}_{62} = (\eta_1 - \eta_2) \sqrt{\eta_L p_{c_L} } \cos \theta_L / 4$ \\
$ {\bf K}_{16} = {\bf K}_{25} = {\bf K}_{52} = {\bf K}_{61} = (\eta_1 - \eta_2) \sqrt{\eta_L p_{c_L} } \sin \theta_L  / 4$ \\
$ {\bf K}_{17} = {\bf K}_{28} = {\bf K}_{71} = {\bf K}_{82} = (\eta_ 2 - \eta_1) \sqrt{\eta_L p_{c_L} \eta_R p_{c_R}} \cos ( \theta_L - \theta_R ) / 4$ \\
$ {\bf K}_{18} = {\bf K}_{81} = - {\bf K}_{27} = - {\bf K}_{72} = (\eta_ 2 - \eta_1) \sqrt{\eta_L p_{c_L} \eta_R p_{c_R}} \sin ( \theta_R - \theta_L ) / 4$ \\
$ {\bf K}_{33} = {\bf K}_{44} = {\bf K}_{55} = {\bf K}_{66} = (\eta_ 2 + \eta_1) / 4 $ \\
$ {\bf K}_{35} = {\bf K}_{53} = {\bf K}_{46} = {\bf K}_{64} = (\eta_ 2 - \eta_1) / 4 $ \\
$ {\bf K}_{37} = {\bf K}_{73} = - {\bf K}_{48} = - {\bf K}_{84} = (\eta_1 - \eta_2) \sqrt{\eta_R p_{c_R} } \cos \theta_R / 4$ \\
$ {\bf K}_{38} = {\bf K}_{47} = {\bf K}_{74} = {\bf K}_{83} = (\eta_1 - \eta_2) \sqrt{\eta_R p_{c_R} } \sin \theta_R / 4$ \\
$ {\bf K}_{68} = {\bf K}_{86} = - {\bf K}_{57} = - {\bf K}_{75} = (\eta_1 + \eta_2) \sqrt{\eta_R p_{c_R} } \cos \theta_R / 4 $ \\
$ {\bf K}_{58} = {\bf K}_{67} = {\bf K}_{76} = {\bf K}_{85} = - (\eta_1 + \eta_2) \sqrt{\eta_R p_{c_R} } \sin \theta_R / 4 $ \\
$ {\bf K}_{77} = {\bf K}_{88} =  (1-p_{c_R}) / 2 + \eta_R p_{c_R} (\eta_1 + \eta_2) /4 $ \\
$ {\bf K}_{12} = {\bf K}_{21} = {\bf K}_{34} = {\bf K}_{43} = {\bf K}_{36} = {\bf K}_{63} = {\bf K}_{45} = {\bf K}_{54} = {\bf K}_{56} = {\bf K}_{65} = {\bf K}_{78} = {\bf K}_{87} = 0  $
		\end{tabular}
		\end{ruledtabular}
\end{table*}
Plugging Eqs.~(\ref{Fid3})--(\ref{Fid4}) into Eq.~(\ref{Fid2}), we finally obtain
\begin{eqnarray}
\label{Fid5}
\lefteqn{\hspace*{-.2in}F_{j,{\rm d}} = \eta_j (1-p_{c_L}) (1-p_{c_R}) \left|   \sqrt{ \eta_L p_{c_L} } d_L^\ast e^{i \theta_L} \right.} \nonumber \\
&\,& \hspace*{-.2in}+ (-1)^j  \left.\sqrt{ \eta_R p_{c_R} } d_R^\ast e^{i \theta_R} \right| ^ 2 / (2 P_j), \;\mbox{ $j = 1,2$.}
\end{eqnarray} 
From Eq.~(\ref{Fid5}), it can be easily seen that the maximum fidelity is achieved by the state given by Eq.~(\ref{optstate}). Also, by assuming $d_L = \pm d_R = 1 / \sqrt{2} $, we find the fidelities of entanglement for the singlet and triplet states as given by Eq.~(\ref{(14)}). Although we only derived Eq.~(\ref{Fid5}) for PNRD systems, one can verify that it also holds for NRPD systems.

The heralding probabilities in Eq.~(\ref{(15)}) can be derived from Eq.~(\ref{(B9)}) by noting that
\begin{equation}
\chi _A^{\rho _{\rm out} } (0 ,0 ,\zeta _{p1} ,\zeta _{p2} ) = (2 \pi) ^2 \sqrt{\det {\bf K'}}\,G({\bm \zeta'}, {\bf K'}) , 
\end{equation}
where
$
{\bm \zeta'} = [\zeta _ {pr} ^ - , \zeta _ {pi} ^ - , \zeta _ {pr} ^ + , \zeta _ {pi} ^ +]^T
$, and
\begin{equation}
{\bf K'} = \frac {1} {\beta_L \beta _R - \delta^2} \left[ \begin{array} {cccc}
\beta_R & 0 & -\delta & 0  \\
0 & \beta_R & 0 & - \delta \\
- \delta & 0 & \beta_L & 0 \\
 0 & -\delta & 0 & \beta_L
 \end{array}
 \right]
 \end{equation}
with $ \sqrt{\det {\bf K'}} = 1 / (\beta_L \beta_R -\delta^2)$. The rest of derivation is straightforward; it parallels what we have done for the fidelities and will be omitted.

\end{document}